\documentclass[%
reprint,
superscriptaddress,
amsmath,amssymb,
aps,
pra,
]{revtex4-2}

\usepackage{graphicx}
\usepackage{dcolumn}
\usepackage{bm}
\usepackage{physics}     
\usepackage{hyperref}
\usepackage{subfigure}
\hypersetup{
	colorlinks = true, 
	linkcolor = blue, 
	urlcolor = blue,
	citecolor = blue,
} 

\begin{document}
	\title{ Distinct Critical Scaling of Quantum Fisher Information in a Quantum Rabi Triangle System}
			\author{Yuyang Tang}
		\affiliation{Ministry of Education Key Laboratory for Nonequilibrium Synthesis 
				and Modulation of Condensed Matter, Shaanxi Province Key Laboratory 
				of Quantum Information and Quantum Optoelectronic Devices, School of 
				Physics, Xi’an Jiaotong University, Xi’an 710049, China}	
		\author{Yu Yang}
		\email{ yangyu1229@hotmail.com}
		\affiliation{Ministry of Education Key Laboratory for Nonequilibrium Synthesis 
				and Modulation of Condensed Matter, Shaanxi Province Key Laboratory 
				of Quantum Information and Quantum Optoelectronic Devices, School of 
				Physics, Xi’an Jiaotong University, Xi’an 710049, China}
		\author{Min An}
		\affiliation{Ministry of Education Key Laboratory for Nonequilibrium Synthesis 
				and Modulation of Condensed Matter, Shaanxi Province Key Laboratory 
				of Quantum Information and Quantum Optoelectronic Devices, School of 
				Physics, Xi’an Jiaotong University, Xi’an 710049, China}
		\author{Fuli Li}
		\affiliation{Ministry of Education Key Laboratory for Nonequilibrium Synthesis 
				and Modulation of Condensed Matter, Shaanxi Province Key Laboratory 
				of Quantum Information and Quantum Optoelectronic Devices, School of 
				Physics, Xi’an Jiaotong University, Xi’an 710049, China}
	
	\date{\today}

	\begin{abstract}
		Critical properties of a quantum system are recognized as valuable resources for quantum metrology. In this work, we investigate the criticality-enhanced sensing in a quantum Rabi triangle system, which exhibits multiple phases. Around the phase boundary, enhanced parameter estimation precision can be achieved by tuning either the scaled coupling strength or the hopping phase controlled by an artificial magnetic field. We observe that the quantum Fisher information shows divergent scaling near different quantum phase transition points, characterized by distinct critical exponents. When the resource consumption is taken into account, we find that the divergent quantum Fisher information can reach the Heisenberg limit. Furthermore, we propose a measurement scheme of the average photon number and the quantum Cramér-Rao bound can be saturated.
	\end{abstract}
	
	\maketitle

	\section{Introduction}

Quantum metrology aims to estimate encoded parameters utilizing quantum systems and quantum resources \cite{RevModPhys.89.035002,demkowicz2015quantum,liu2020quantum}, such as quantum entanglement and quantum control \cite{PhysRevLett.96.010401,giovannetti2011advances,PhysRevLett.117.160801,pang2017optimal,hou2021zero}, while how to achieve high-precision quantum sensing that outperforms classical strategies has become a central focus of recent research. Recently, it has been discovered that near quantum phase transition (QPT) points, sensing precision can be greatly enhanced beyond the standard quantum limit \cite{PhysRevA.93.022103,RevModPhys.90.035006,ding2022enhanced,PhysRevLett.132.060801,PhysRevA.109.022604}. The essential mechanism is the energy gap closing at the phase transition point, where even a slight variation in the Hamiltonian can induce a significant change in the observable quantity \cite{PhysRevB.76.180403,PRXQuantum.3.010354}. This behavior manifests as the divergence of the quantum Fisher information (QFI), indicating the potential for achieving arbitrarily high precision of quantum sensing.

The critical property has been identified as an important metrological resource in a wide range of quantum systems that exhibit QPTs, such as the quantum Rabi model (QRM) \cite{PhysRevLett.126.010502,PhysRevA.110.012413,PhysRevA.110.022611}, the Lipkin-Meshkov-Glick (LMG) model and the XY spin-chain model \cite{PhysRevA.80.012318,PhysRevA.90.022111,CAROLLO20201,PhysRevLett.126.200501}. However, in such systems, the critical resources are predominantly localized near phase transition points \cite{PhysRevLett.133.120601,PhysRevLett.134.190802}. Owing to the restriction of only a single control parameter, it becomes challenging to fully harness these resources. Therefore, introducing additional controllable parameters is essential to enhance the tunability of the system and to access richer critical resources.

Compared with the conventional QRM, the quantum Rabi triangle (QRT) model introduces an artificial magnetic field that gives rise to a chiral superradiant phase with broken time-reversal symmetry \cite{PhysRevLett.127.063602,PhysRevLett.129.183602}. In the fast-oscillator limit, which effectively plays the role of a thermodynamic limit \cite{PhysRevA.85.043821,hwang2015quantum}, the QRT system exhibits three different phases: the normal phase (NP), the ferromagnetic superradiant phase (FSP), and the chiral superradiant phase (CSP). When the system undergoes a transition from the NP to either of the superradiant phases, a second-order quantum phase transition occurs, whereas the transition between the two superradiant phases is of first-order. The coexistence of these multiple phases provides access to richer critical resources, making the QRT model an ideal platform for exploring critical quantum sensing. 
 
 In this work, we propose a circuit quantum electrodynamics (QED) implementation of the QRT model and evaluate the QFI across its three phases around the phase boundaries. We find that the QFI is enhanced by both first-order and second-order phase transitions. It exhibits a divergent behavior around the boundary of the second-order phase transition and the triple point. By further analyzing the scaling behavior of the divergent QFI, we observe that, in the presence of an artificial magnetic field ($\theta \neq 0$), the Heisenberg limit (HL) $\sim \langle N\rangle^2 T^2$ can be achieved when the system is tuned to its critical points, either through the scaled coupling strength or the hopping phase induced by the artificial magnetic field. Furthermore, we propose a feasible measurement scheme based on the average photon number and demonstrate that the QFI can saturate the quantum Cramér-Rao bound (QCRB). 

The remainder of this paper is organized as follows. In section \ref{section:2}, We first introduce the QRT Hamiltonian and then present its implementation within a circuit QED scheme, followed by a discussion of the Hamiltonian transformations in three different phases. We calculate the QFI and demonstrate its enhancement around the phase boundary in section \ref{section:3}. Then we discuss the scaling behavior of the QFI in section \ref{section:4}. In section \ref{section:5}, we investigate the measurement scheme of the average photon number. Finally, we give a conclusion in section \ref{section:6}.

\section{Model}	\label{section:2}

The QRT model describes a system consisting of three cavities, each interacting with a two-level atom, with photon hopping between neighboring cavities \cite{PhysRevLett.127.063602,PhysRevA.110.013713}. The system Hamiltonian reads
\begin{equation}\label{eq1}
	H_{\mathrm{QRT}}=\sum_{n=1}^3H_{R,n}+\sum_{n=1}^3J(e^{i\theta}a_n^\dagger a_{n+1}+e^{-i\theta}a_{n+1}^\dagger a_n),
\end{equation}
with $a^{\dagger}_n\ (a_n)$ the bosonic creation (annihilation) operator of the photon in $n$th cavity, $J$ the hopping amplitude between cavities $n$ and $n+1$, and $e^{\pm i\theta}$ the phase arising from an artificial magnetic field. The Hamiltonian of the $n$th cavity takes the form of the quantum Rabi model (QRM):
\begin{equation}\label{eq2}
	H_{R,n}=\omega a^{\dagger}_na_n+\frac{\Delta}{2}\sigma_n^z+g(a^{\dagger}_n+a_n)\sigma_n^x.
\end{equation}
Here $\omega$ is the frequency of photon, $\Delta$ is the energy gap between two internal states of the atom, $g$ is the cavity-atom coupling strength, and $\hat{\sigma}^{x,y,z}_n$ represent the Pauli matrices of the two-level atom which is coupled to the $n$th cavity.

\begin{figure}[h!]
	\centering\includegraphics[width=0.4\textwidth]{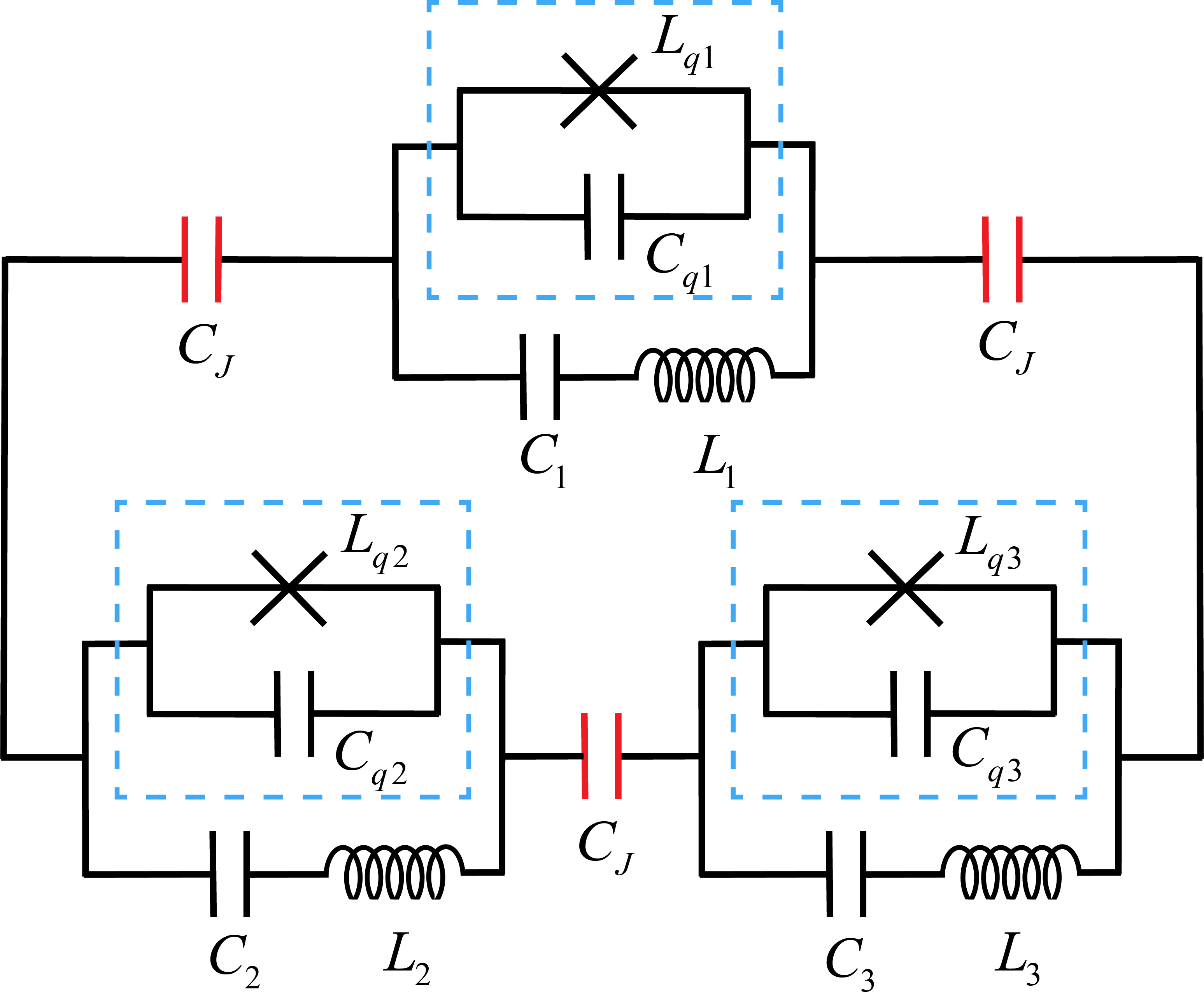}
	\caption{Circuit diagram of a three-mode superconducting loop: each anharmonic LC oscillator (dashed boxes) consists of a Josephson junction ($L_{q1},\ L_{q2},\ L_{q3}$) and a capacitance ($C_{q1},\ C_{q2},\ C_{q3}$), which is coupled to the electromagnetic cavity mode of LC resonator ($L_{1},\ L_{2},\ L_{3},\ C_{1},\ C_{2},\ C_{3}$). The capacitance $C_J$ is added between cavities to modulate the hopping strength.}\label{Fig:fig1}
\end{figure}

Such a Hamiltonian can be realized in a circuit QED setup, as illustrated in Fig.~\ref{Fig:fig1}. Each "artificial atom", consisting of a Josephson junction shunted by a capacitance, is individually coupled to an electromagnetic cavity mode formed by a series LC resonator. The hopping strength between adjacent cavities is tunable via the coupling capacitance $C_J$ \cite{manucharyan2017resilience,PhysRevB.95.245115,PhysRevA.97.013851,PhysRevResearch.1.033128}.

Similar to the QRM, this QRT system undergoes a second-order phase transition from the NP to the superradiant phase (SP), when the modified cavity-atom coupling strength $g_1=g/\sqrt{\Delta\omega}$ is approached the critical point $g_{1c}$. The corresponding order parameter is given by $\langle a_N\rangle$, representing the mean-field amplitude of the $n$th cavity, which continuously increases from zero.

Besides, during the modulation of the hopping phase $\theta$, complex hopping amplitudes emerge, leading to the breaking of time-reversal symmetry when $\theta\neq m\pi\ (m\in\mathbb{Z})$. It gives rise to the emergence of two distinct superradiant phases: the FSP and the CSP. The first-order phase transition between these two phases can be characterized by a discontinuous change in the order parameter $\langle I_{\mathrm{ph}}\rangle$, where $I_{\mathrm{ph}}=i[(a_1^{\dagger}a_2+a_2^{\dagger}a_3+a_3^{\dagger}a_1)-\mathrm{H.c.}]$ represents the photon current induced by the artificial magnetic field \cite{PhysRevLett.127.063602}. 

These three phases are discussed in detail below, as shown in Fig. \ref{Fig:fig2}(a).

\subsection{Normal Phase}

 In the limit of $\Delta/\omega\rightarrow\infty$, we apply the Schrieffer-Wolff (SW) transformation on the Hamiltonian (\ref{eq1}) (See Appendix \ref{appendix1}). Then the effective Hamiltonian in the NP is given by
\begin{eqnarray}\label{eq5}
	H_{\mathrm{NP}}&=&\sum_{n=1}^{3}\omega a_{n}^{\dagger}a_{n}-g_{1}^{2}\omega\left(a_{n}^{\dagger}+a_{n}\right)^{2}\nonumber\\
	&\ &+Ja_{n}^{\dagger}(e^{i\theta}a_{n+1}+e^{-i\theta}a_{n-1})-\frac{3\Delta}{2}.
\end{eqnarray}

By employing the Fourier transformation $a_n^{\dagger}=\frac{1}{\sqrt{3}}\sum_qe^{inq}a_q^{\dagger}$, the Hamiltonian (\ref{eq5}) is transformed to 
\begin{equation}\label{eq6}
	H_{\mathrm{NP}}=\sum_q\omega_qa_q^\dagger a_q-g_1^2\omega(a_qa_{-q}+a_q^\dagger a_{-q}^\dagger) +E_{\mathrm{np}},
\end{equation}
where $\omega_q=\omega-2g_1^2\omega+2J\cos(\theta-q),\ q=0,\pm2\pi/3,\ E_{\mathrm{np}}=-3\Delta/2-3g_1^2\omega$. The Hamiltonian can be diagonalized using the Bogoliubov transformation (see Appendix \ref{appendix2}). The diagonalized Hamiltonian is
\begin{equation}
	H_{\mathrm{NP}}=\sum_q\epsilon_q b_q^{\dagger}b_q+E_q,
\end{equation}
with
\begin{eqnarray}
	\epsilon_q&=&\frac{1}{2}(\sqrt{(\omega_q+\omega_{-q})^2-16\omega^2g_1^4}+\omega_q-\omega_{-q}),\nonumber\\
	E_q&=&\frac{1}{4} ( \sqrt{(\omega_q+\omega_{-q})^2-16\omega^2g_1^4}  -(\omega_{q} + \omega_{-q})).
\end{eqnarray}

\subsection{Superradiant Phase}

\begin{figure}[h!]
	\centering\includegraphics[width=0.5\textwidth]{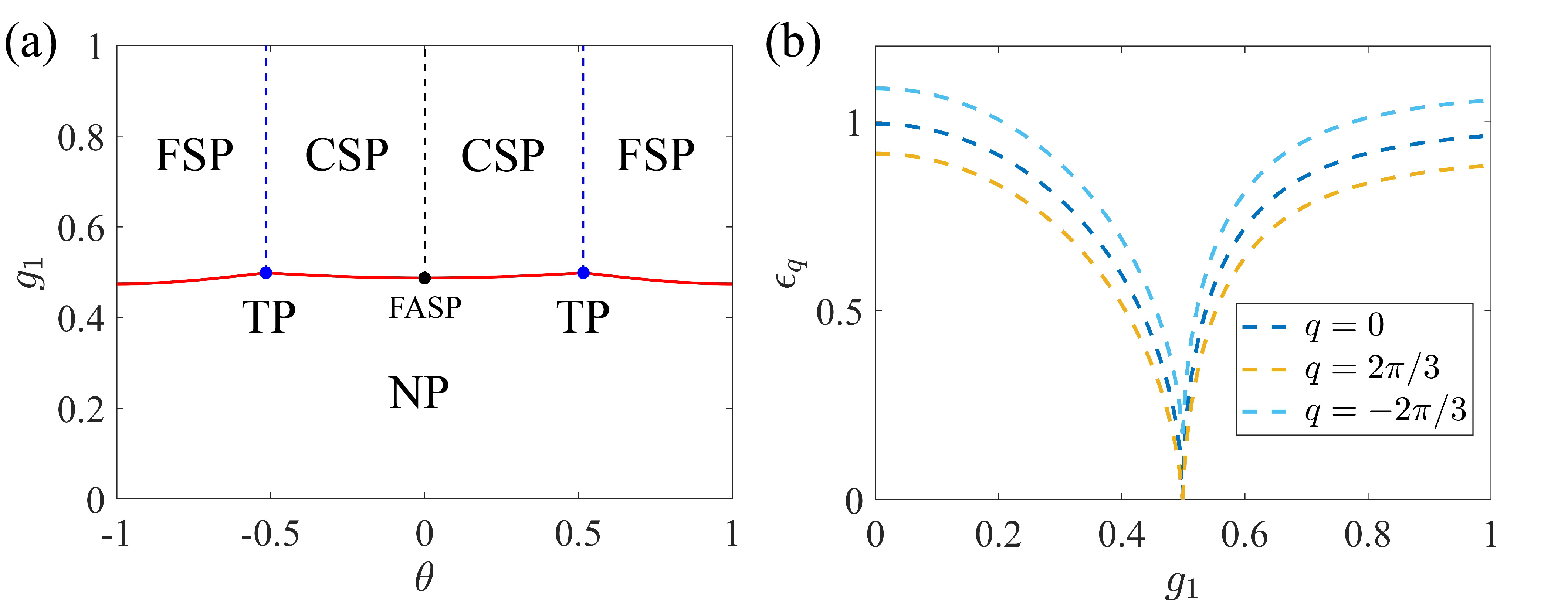}
	\caption{(a)The phase diagram of the quantum Rabi triangle model. The red solid line stands for the second-order phase transition boundary $g_{1c}(q,\theta)$ between the normal phase (NP) and the superradiant phases. The blue dashed line $\pm\theta_c$ stands for the first-order phase transition boundary between the ferromagnetic superradiant phase (FSP) and the chiral superradiant phase (CSP). The two lines cross at the triple point (TP). The black dashed line stands for the special frustrated antiferromagnetic superradiant phase (FASP). (b)The excitation spectrum $\epsilon_q$ as a function of $g_1$ with $\theta=-\theta_c$ for $q={0,\pm2\pi/3}$.}\label{Fig:fig2}
\end{figure}

When $g_1$ is approached the critical point $g_{1c}$, the excitation spectrum $\epsilon_q$ tends to zero, as shown in Fig. \ref{Fig:fig2}(b). When the system undergoes a second-order phase transition from the NP to the SP, a nonzero mean-field amplitude $\langle a_N\rangle$ emerges. Owing to this, the creation and annihilation operators are displaced as
\begin{equation}\label{eq7}
	a_n^\dagger\to\tilde{a}_n^\dagger+\alpha_n^*, a_n\to\tilde{a}_n+\alpha_n,
\end{equation}
with the displacement $\alpha_n=A_n+iB_n$ (see Appendix \ref{appendix2}). 

The effective Hamiltonian after the SW transformation is 
\begin{eqnarray}\label{eq10}
	H_{\mathrm{SP}}&=&\sum_{n=1}^3\omega\tilde{a}_n^\dagger\tilde{a}_n-\frac{\lambda_n^2}{\Delta_n}\left(\tilde{a}_n^{\dagger}+\tilde{a}_n\right)^2\nonumber\\
	&\ &+J\tilde{a}_n^\dagger(e^{i\theta}\tilde{a}_{n+1}+e^{-i\theta}\tilde{a}_{n-1})-\frac{\Delta_n}{2}+E_0,
\end{eqnarray}
with the rescaled frequency $\Delta_n=\sqrt{\Delta^2+16g^2A_n^2}$, the effective coupling strength $\lambda_n=g\Delta/\Delta_n$, the constant term $E_0=\sum_{n=1}^{3}\omega\alpha_{n}^{*}\alpha_{n}+J\alpha_{n}^{*}(e^{i\theta}\alpha_{n+1}+e^{-i\theta}\alpha_{n-1})$ (See Appendix \ref{appendix1}).

In the FSP, the displacement $\alpha_n$ for each cavity is real and identical, therefore we set $\lambda_n=\lambda^{\prime}$ and $\Delta_n=\Delta^{\prime}$. The Hamiltonian after Fourier transformation is 
\begin{equation}\label{eq11}
	H_{\mathrm{FSP}}= \sum_q\omega_q^{\prime}\tilde{a}_q^\dagger \tilde{a}_q-\frac{\lambda^{\prime 2}}{\Delta^{\prime}}(\tilde{a}_q\tilde{a}_{-q}+\tilde{a}_q^\dagger \tilde{a}_{-q}^\dagger)+E_0-\frac{3\Delta^{\prime}}{2}.
\end{equation}
Similarly, after the Bogoliubov transformation (see Appendix \ref{appendix2}), the diagonalized Hamiltonian takes the form
\begin{equation}
	H_{\mathrm{FSP}}=\sum_q\epsilon_q^{\prime} \tilde{b}_q^{\dagger}\tilde{b}_q+E_q^{\prime},
\end{equation}
with
\begin{eqnarray}
	\epsilon_q^{\prime}&=&\frac{1}{2}\left(\omega_q^{\prime}-\omega_{-q}^{\prime}+\sqrt{(\omega_q^{\prime}+\omega_{-q}^{\prime})^2-16(\frac{\lambda^{\prime 2}}{\Delta^{\prime}})^2}\right),\nonumber\\
	E_{q}^{\prime}&=&-\frac{\lambda^{'2}}{\Delta^{\prime}}+\frac{1}{2}\sum_{q}(\epsilon_{q}^{\prime}-\omega_{q}^{\prime}),
\end{eqnarray}
where $\tilde{b}_q^{\dagger}(\tilde{b}_q)$ is the transformed creation(annihilation) operator, $\omega_{q}^{\prime}=\omega-2\frac{\lambda^{\prime 2}}{\Delta^{\prime}}+2J\cos(\theta-q)$.

By modulating the artificial magnetic field, the change of $\theta$ makes the $\alpha_n$ complex in the CSP. In such phase, $\Delta_n\neq\Delta_{n^{\prime}}$, the Hamiltonian (\ref{eq10}) cannot be transformed into $q$ space. By directly applying Bogoliubov transformation to the Hamiltonian (\ref{eq10}), the diagonalized Hamiltonian has the form
\begin{equation}
	H_{\mathrm{CSP}}=2\sum_n^3\epsilon_n\tilde{c}^{\dagger}_n\tilde{c}_n+\frac{\epsilon_n-\omega}{2},
\end{equation}
where $\tilde{a}_n^{\dagger}=\sum_{i=1}^3T_{n,i}\tilde{c}_i+T_{n,i+3}\tilde{c}_i^{\dagger}$, $T$ is the transform matrix of the diagonalization, and $\epsilon_n$ is also obtained by the diagonalization (see Appendix \ref{appendix2}).

The phase boundary between the NP and the SP is given by
\begin{equation}
	g_{1c}(q,\theta)=\sqrt{\frac{{(2 J \cos (q-\theta )+\omega)(2 J \cos (q +\theta)+\omega) }}{{4 J\omega \cos (q-\theta )+4 J\omega \cos (q +\theta)+4 \omega^2 }}},
\end{equation}
which depends on $q$ and $\theta$, taking different values for different phases. The three phases coexist at the triple point $(g_{1c}(q,\pm\theta_c),\pm\theta_c)$, as shown in Fig. \ref{Fig:fig2}(a). The critical hopping phase $\theta_c$ is determined by the condition $g_{1c}(q,\theta)=g_{1c}(q',\theta)$, where $q$ and $q'$ correspond to different phases, yielding
\begin{equation}
	\theta_c=\cos ^{-1}\left(-\frac{2 J}{\sqrt{8 J^2+\omega ^2}+\omega }\right).
\end{equation}
In particular, at $\theta=0$, the chiral characteristic vanishes, and the system exhibits a frustrated antiferromagnetic order, known as the frustrated antiferromagnetic superradiant phase (FASP)\cite{PhysRevLett.129.183602, PhysRevA.110.013713}.

\section{The enhanced quantum Fisher information around the phase boundary} \label{section:3}

Quantum sensing is devoted to estimate unknown parameters with high precision by exploiting unique quantum properties, and it can be systematically analyzed using the quantum parameter estimation theory \cite{polino2020photonic}. In single-parameter quantum estimation, the achievable precision is fundamentally limited by the QCRB, which is governed by the QFI. In this section, we explore the enhanced QFI of the parameter $g_1$ around the phase boundary. 

For a pure state, the QFI with respect to the parameter $\lambda$ can be expressed as \cite{GIBBONS1992147,PhysRevLett.72.3439,PhysRevLett.99.100603}
\begin{eqnarray}\label{eq15}
	I(\lambda)&=& 4\left(\left\langle {\frac{{d\psi_0(\lambda) }}{{d\lambda}}}\bigg| {\frac{{d\psi_0(\lambda) }}{{d\lambda}}} \right\rangle  -\bigg |\left\langle \psi_0(\lambda) \bigg | {\frac{{d\psi_0(\lambda) }}{{d\lambda}}} \right\rangle \bigg|^2 \right),\nonumber\\
\end{eqnarray}
where $\psi_0(\lambda)$ is the encoded ground state that contains the information of the parameter $\lambda$. By the first order perturbative expansion \cite{sakurai2020modern}
\begin{equation}\label{eq16}
	\ket{\psi(\lambda+\delta\lambda)}\sim\ket{\psi(\lambda)}+\sum_{n\neq0}\frac{\bra{\psi_n(\lambda)}\delta H\ket{\psi_0(\lambda)}}{E_0(\lambda)-E_n(\lambda)}\ket{\psi_n(\lambda)},
\end{equation}
where $H$ is the system Hamiltonian and $\delta H=H(\lambda+\delta\lambda)-H(\lambda)$. 

By substituting Eq. (\ref{eq16}) into Eq. (\ref{eq15}), we obtain the QFI as
\begin{equation}\label{eq17}
	I(\lambda)=4\sum_{n\neq0}\frac{|\langle\psi_n(\lambda)|\partial_{\lambda} H|\psi_0(\lambda)\rangle|^2}{[E_n(\lambda)-E_0(\lambda)]^2}.
\end{equation}
It follows from the above equation that the closing of the energy gap directly results in the divergence of the QFI. Specifically, in the QRT model, when $g_1$ is approached the triple point $g_{1c}(0,-\theta_c)$, the energy gap closing is shown in Fig. \ref{Fig:fig2}(b). This divergent behavior allows us to further evaluate the QFI around the phase boundaries $g_{1c}(q,\theta)$ in the QRT model.

\begin{widetext}
	
In the NP, the partial derivative of Hamiltonian (\ref{eq6}) with respect to $g_1$ is
\begin{eqnarray}\label{eq18}
	\partial_{g_1}H_{\mathrm{NP}}&=&\sum_q-2g_1\omega(\mu_q+\nu_q)^2[(b_qb_{-q}+b_q^{\dagger}b_{-q}^{\dagger})+2b_q^{\dagger}b_{q}+1],
\end{eqnarray}
with the transformed bosonic modes $\{b_q,\ b_q^{\dagger}\}$. Since we have applied the Bogoliubov transformation, the corresponding ground state should be transformed to the squeezed vacuum state (see Appendix \ref{appendix3}). Substituting Eq. (\ref{eq18}) into Eq. (\ref{eq17}) yields the expression of the QFI
\begin{eqnarray}\label{eq19}
	I_{\mathrm{NP}}(g_1)&=&{16\omega^2g_1^2}\sum_q\frac{2}{(\omega_q+\omega_{-q}-4\omega g_1^2)^2}.
\end{eqnarray}
	
Similarly, in the FSP, the partial derivative of Hamiltonian (\ref{eq11}) with respect to $g_1$ is
\begin{eqnarray}\label{eq35}
	\partial_{g_1}H_{\mathrm{FSP}}&=&\sum_q-\partial_{g_1}(\frac{\lambda^{\prime 2}}{\Delta^{\prime}})[(\mu_q+\nu_q)^2(\tilde{b}_q\tilde{b}_{-q}+\tilde{b}_q^{\dagger}\tilde{b}_{-q}^{\dagger})+2(\mu_q+\nu_q)^2\tilde{b}_q^{\dagger}\tilde{b}_{q}]-\partial_{g_1}(\frac{\lambda^{\prime 2}}{\Delta^{\prime}})(\mu_q+\nu_q)^2+\partial_{g_1}(E_0-\frac{3\Delta^{\prime}}{2}).\nonumber\\
\end{eqnarray}
with the transformed bosonic modes $\{\tilde{b}_q,\ \tilde{b}_q^{\dagger}\}$. The explicit form of the QFI is obtained by inserting Eq. (\ref{eq35}) into Eq. (\ref{eq17})
\begin{eqnarray}\label{eq20}
	I_{\mathrm{FSP}}(g_1)
	&=&\frac{(2 J \cos \theta +\omega )^6}{64 g_1^{10} \omega ^4}\sum_q\frac{2}{(\omega_q^{\prime}+\omega_{-q}^{\prime}-4{\lambda^{'2}}/{\Delta^{\prime}})^2}.
\end{eqnarray}
	
In the CSP, the partial derivative of Hamiltonian (\ref{eq10}) with respect to $g_1$ is 
\begin{eqnarray}\label{eq36}
	\partial_{g_1}H_{\mathrm{CSP}} &=&\sum_{n=1}^3\partial_{g_1}(\frac{\lambda^{2}_n}{\Delta_{n}})\sum_{i=1}^{3} \sum_{j=1}^{3}[(T_{n,i} + T_{n,i+3}^*)(T_{n,j} + T_{n,j+3}^*) \tilde{c}_i \tilde{c}_j + (T_{n,i} + T_{n,i+3}^*)(T_{n,j+3} + T_{n,j}^*) \tilde{c}_i \tilde{c}_j^{\dagger} \nonumber\\
	&\ &+ (T_{n,i+3} + T_{n,i}^*)(T_{n,j} + T_{n,j+3}^*) \tilde{c}_i^{\dagger} \tilde{c}_j+ (T_{n,i+3} + T_{n,i}^*)(T_{n,j+3} + T_{n,j}^*) \tilde{c}_i^{\dagger} \tilde{c}_j^{\dagger}]+\partial_{g_1}(E_0-\sum_{n=1}^3\frac{\Delta_n}{2}).
\end{eqnarray}
with the transformed bosonic modes $\{\tilde{c}_i,\ \tilde{c}_i^{\dagger}\}$, and $T_{n,i}$ denotes the matrix element of the transformation matrix $T$ used for the Hamiltonian diagonalization. By substituting Eq. (\ref{eq36}) into Eq. (\ref{eq17}), we obtain the explicit expression for the QFI
\begin{eqnarray}\label{eq21}
	I_{\mathrm{CSP}}(g_1)
	&=&4\sum_{i=1}^{3}(\frac{2(\sum_{n=1}^3\partial_{g_1}(\frac{\lambda^{2}_n}{\Delta_{n}})(T_{n,i} + T_{n,i+3}^*)^2)(\sum_{m=1}^3\partial_{g_1}(\frac{\lambda^{2}_m}{\Delta_{m}})(T_{m,i+3} + T_{m,i}^*)^2)}{4\epsilon_i^2}\nonumber\\
	&\ &+4\sum_{j=1(j\neq i)}^3\frac{(2\sum_{n=1}^3\partial_{g_1}(\frac{\lambda^{2}_n}{\Delta_{n}})(T_{n,i} + T_{n,i+3}^*)(T_{n,j} + T_{n,j+3}^*))(2\sum_{m=1}^3\partial_{g_1}(\frac{\lambda^{2}_m}{\Delta_{m}})(T_{m,i+3} + T_{m,i}^*)(T_{m,j+3} + T_{m,j}^*))}{(\epsilon_i+\epsilon_j)^2}).\nonumber\\
\end{eqnarray}

\end{widetext}

\begin{widetext}

\begin{figure}[h!]
	\centering\includegraphics[width=0.95\textwidth]{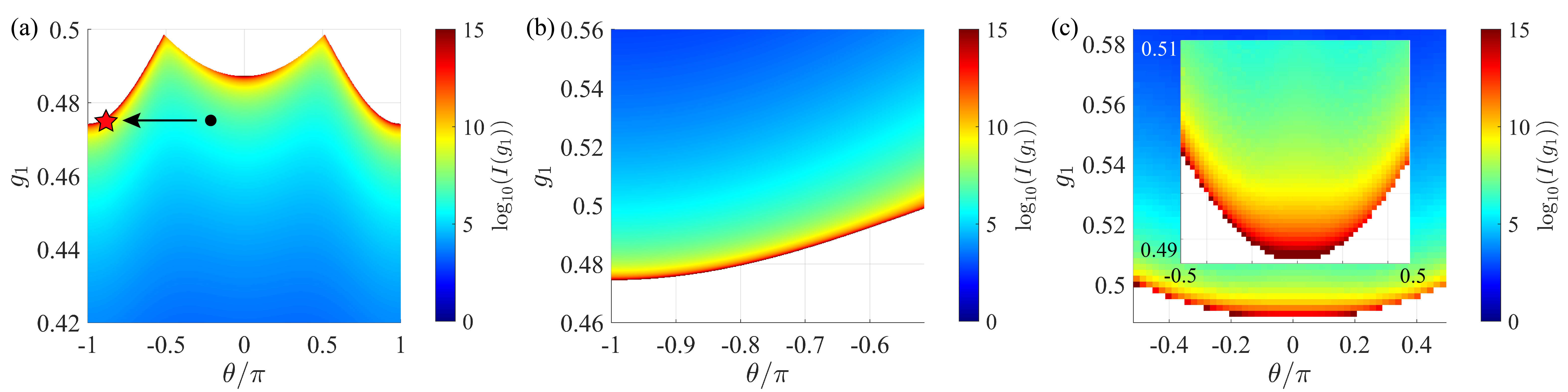}
	\caption{The QFI as a function of $g_1$ and $\theta$ normalized by $\pi$. (a) $I_{\mathrm{NP}}(g_1)$ in the NP. The black dot $(g_{1c}(-2\pi/3,\theta_0),\theta_0)$ is away from the phase boundary, and the red star $(g_{1c}(0,\theta^{\prime}),\theta^{\prime})$ is near the phase boundary. (b) $I_{\mathrm{FSP}}(g_1)$ in the FSP. (c) $I_{\mathrm{CSP}}(g_1)$ in the CSP. The inset presents a zoomed-in view of $I_{\mathrm{CSP}}(g_1)$ for $g_1$ in range of 0.49 to 0.51.}\label{Fig:fig3}
\end{figure}

\end{widetext}

In the NP, Eq. (\ref{eq19}) indicates that as $g_1$ is approached $g_{1c}$, the denominator tends to zero, leading to a divergence of the QFI $I_{\mathrm{NP}}(g_1)$. As shown in Fig. \ref{Fig:fig3}(a), this manifests as an enhancement of $I_{\mathrm{NP}}(g_1)$ around the phase boundary $g_{1c}(q,\theta)$. Similarly, in the FSP, Eq. (\ref{eq20}) predicts that $I_{\mathrm{FSP}}(g_1)$ diverges when $g_1$ is approached $g_{1c}$, which is reflected in an enhancement of $I_{\mathrm{FSP}}(g_1)$ around the phase boundary  $g_{1c}(q,\theta)$, as shown in Fig. \ref{Fig:fig3}(b). In the CSP, according to Eq. (\ref{eq21}), the lowest excitation energy $\epsilon_1$ tends to zero as $g_1$ is approached $g_{1c}$, resulting in a divergence of $I_{\mathrm{CSP}}(g_1)$. Consistently, $I_{\mathrm{CSP}}(g_1)$ exhibits a significant enhancement around the phase boundary $g_{1c}(q,\theta)$, as shown in Fig. \ref{Fig:fig3}(c).

It shows that in each phase, the QFI exhibits a pronounced enhancement around the phase boundary $g_{1c}(q,\theta)$. The phase boundary $g_{1c}(q,\theta)$ depends on $q$ and $\theta$, indicating that besides the scaled coupling strength $g_{1}$, the hopping phase $\theta$ tuned by the artificial magnetic field can also be a controllable parameter. For instance, as shown in Fig. \ref{Fig:fig3}(a), when the system is initially prepared in the NP at the black dot (away from the phase boundary $g_{1c}(-2\pi/3,\theta_0)$), varying $\theta_0$ to a new value $\theta^{\prime}$ through modulation of the artificial magnetic field can drive the system closer to the phase boundary $g_{1c}(0,\theta^{\prime})$, as indicated by the red star.

\section{The scaling of quantum Fisher information near the phase transition point} \label{section:4}

In a system undergoing a phase transition controlled by a parameter $\lambda$, the correlation length generally diverges near the critical point $\lambda_c$ as $\xi\sim|\lambda-\lambda_c|^{-\nu}$, where $\nu$ is the critical exponent \cite{fisher1967theory,RevModPhys.49.435}. This critical scaling behavior is also manifested in the divergence of the QFI.

\begin{widetext}
	
	\begin{figure}[h!]
		\centering\includegraphics[width=0.9\textwidth]{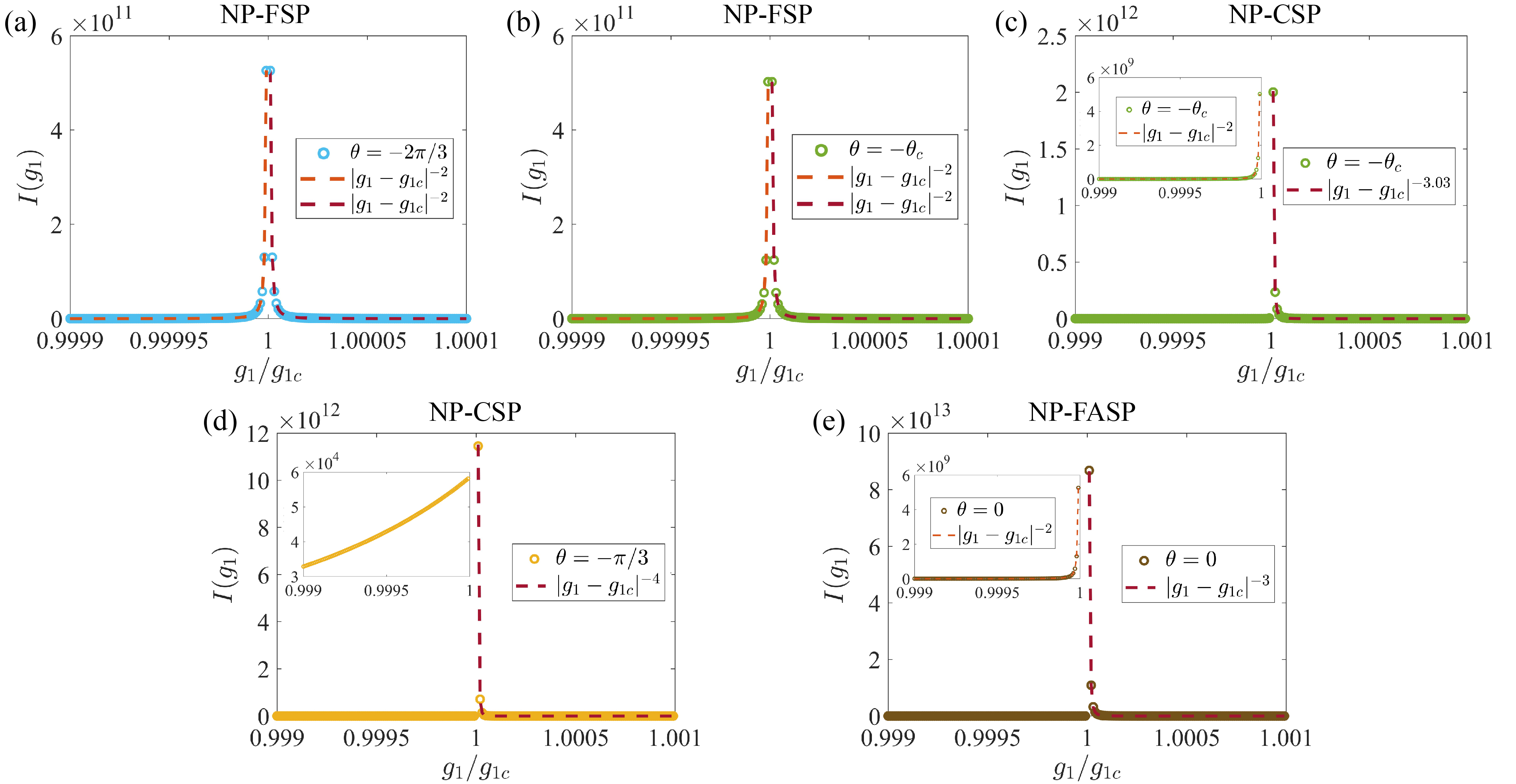}
		\caption{The scaling of the QFI $I(g_1)$ as a function of $g_1$ normalized by $g_{1c}$ with different $\theta$. (a) Scaling behavior of $I(g_1)$ near the critical point $(g_{1c}(0,-2\pi/3),-2\pi/3)$ for the NP–FSP transition. (b) Scaling behavior of $I(g_1)$ near the triple point $(g_{1c}(0,-\theta_c),-\theta_c)$ for the NP-FSP transition. (c) Scaling behavior of $I(g_1)$ near the triple point $(g_{1c}(0,-\theta_c),-\theta_c)$ for the NP-CSP transition. The inset presents a zoomed-in view of $I_{\mathrm{NP}}(g_1)$. (d) Scaling behavior of $I(g_1)$ near the critical point $(g_{1c}(-2\pi/3,-\pi/3),-\pi/3)$ for the NP-CSP transition. The inset presents a zoomed-in view of $I_{\mathrm{NP}}(g_1)$. (e) Scaling behavior of $I(g_1)$ near the critical point $(g_{1c}(-2\pi/3,0),0)$ for the NP-FASP transition. The inset presents a zoomed-in view of $I_{\mathrm{NP}}(g_1)$.} \label{Fig:fig4}
	\end{figure}
	
\end{widetext}

For four fixed values of $\theta$, the corresponding behavior of $I(g_1)$ as a function of the scaled coupling strength $g_1$  normalized by $g_{1c}$ near the critical point $g_{1c}(q,\theta)$ is presented in Fig. \ref{Fig:fig4}. As shown in Fig. \ref{Fig:fig4}(a), when the phase transition occurs from the NP to the FSP, the QFI diverges as $|g_1 - g_{1c}|^{-2}$, characterized by a critical exponent of 2.
At the triple point $\theta = -\theta_c$, the QFI exhibits distinct divergent behaviors on the two sides of $-\theta_c$, corresponding to transitions from the NP to the FSP and from the NP to the CSP. Specifically, as shown in Figs. \ref{Fig:fig4}(b) and \ref{Fig:fig4}(c), the critical exponent remains 2 in the NP and the FSP but increases to approximately 3 in the CSP. For $\theta=-\pi/3$, the QFI diverges only in the CSP, with a critical exponent of 4, as shown in Fig. \ref{Fig:fig4}(d). Finally, as shown in Fig. \ref{Fig:fig4}(e), during the transition from the NP to the FASP, the QFI diverges with a critical exponent of 2 in the NP and 3 in the FASP.

In quantum sensing, enhancing precision beyond the standard quantum limit (SQL) toward the HL inevitably involves the consumption of quantum resources \cite{giovannetti2004quantum,PhysRevLett.98.090401}. In our case, the average photon number $\langle N\rangle$ of the cavity, which exhibits the most rapid increase near the phase point is taken as the resource, and $T$ is the evolution time. Consequently, the SQL scales as $\langle N\rangle T$, while the HL scales as $\langle N\rangle^{2}T^{2}$ \cite{PRXQuantum.3.010354,PhysRevLett.126.070503}. 

In the adiabatic protocol, the ground-state preparation time scales as $T\sim1/\epsilon$ , with $\epsilon$ the first excitation energy (see Appendix \ref{appendix4}) \cite{PhysRevLett.124.120504,alushi2025collective}. In the NP and the FSP, the average photon number  of the first cavity $\langle N_1\rangle$ scales as $|g_1-g_{1c}|^{-1/2}$ and the first excitation energy $\epsilon_0$ scales as $|g_1-g_{1c}|^{1/2}$ (see Appendix \ref{appendix5}). It indicates that the HL can be achieved in the NP and the FSP with $I(g_1)\sim\langle N_1\rangle^{2}T^{2}\sim|g_1-g_{1c}|^{-2}$. In the CSP with $\theta=-\theta_c$, $\langle N_1\rangle$ scales as $|g_1-g_{1c}|^{-1/2}$ and the first excitation energy $\epsilon_1$ scales as $|g_1-g_{1c}|^{1}$, while $I(g_1)\sim|g_1-g_{1c}|^{-3}$ also achieves the HL. For $\theta=-\pi/3$, $\langle N_1\rangle$ scales as $|g_1-g_{1c}|^{-1/2}$ and the first excitation energy $\epsilon_1$ scales as $|g_1-g_{1c}|^{3/2}$. $I(g_1)\sim|g_1-g_{1c}|^{-4}$ still obeys the HL. However, in the FASP with $\theta=0$, $\langle N_1\rangle$ scales as $|g_1-g_{1c}|^{-1}$ and the first excitation energy $\epsilon_1$ scales as $|g_1-g_{1c}|^{1}$, therefore $I(g_1)\sim|g_1-g_{1c}|^{-3}$ obeys the sub-Heisenberg scaling.

\begin{widetext}
	
	\begin{figure}[h!]
		\centering\includegraphics[width=\textwidth]{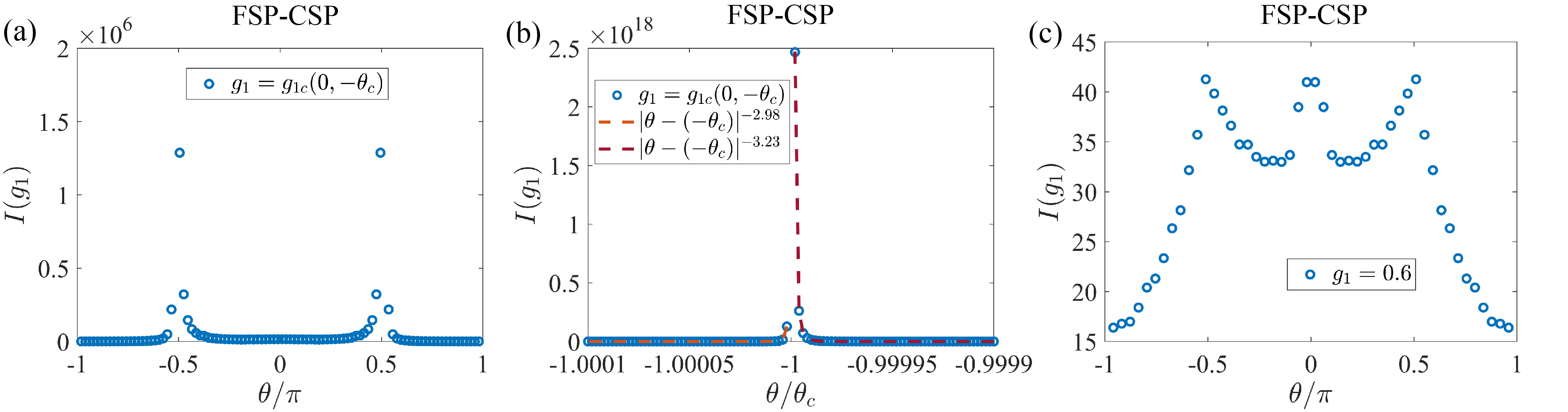}
		\caption{The scaling of the QFI $I(g_1)$ as a function of $\theta$ normalized by $\pi$ (or $\theta_c$) with different $g_1$. (a) The divergent feature of $I(g_1)$ around the triple points $(g_{1c}(0,-\theta_c),\pm\theta_c)$. (b) A  zoomed-in view of $I(g_1)$ around the triple point $(g_{1c}(0,-\theta_c),-\theta_c)$, illustrating the scaling behavior of $I(g_1)$ for the FSP–CSP transition. (c) The peak feature of $I(g_1)$ around the phase transition points $(0.6,\pm\theta_c)$ and $(0.6,0)$ with $g_1$ far from the phase boundary $g_{1c}$.} \label{Fig:fig5}
	\end{figure}
	
\end{widetext}

The QFI $I(g_1)$ as a function of $\theta$  normalized by $\theta_{c}$ is shown in Fig. \ref{Fig:fig5}. For fixed values of $g_1$, the system undergoes a first-order phase transition from the FSP to the CSP. We consider two scenarios: one where $g_1$ crosses the phase boundary $g_{1c}(q,\theta)$ near the triple point $(g_{1c}(0,-\theta_c),-\theta_c)$, and another where far from the phase boundary $g_{1c}(q,\theta)$. In Fig. \ref{Fig:fig5} (a) and (b), it reveals that $I(g_1)$ diverges in the vicinity of the triple point, scaling approximately as $|\theta-(-\theta_c)|^{-3}$. It obeys the HL around the triple point, where  $\langle N_1\rangle$ scales as $|\theta-(-\theta_c)|^{-1/2}$ and the first excitation energy $\epsilon_1$ scales as $|\theta-(-\theta_c)|^{1}$ (see Appendix \ref{appendix5}). In contrast, when $g_1=0.6$, the system is far from the phase boundary $g_{1c}(q,\theta)$. In this regime, only the first order phase transition occurs as $\theta$ is approached $\pm\theta_c$, without any energy gap closing. Consequently, the QFI exhibits only finite peaks near the phase transition points $(0.6,\ \pm\theta_c)$ and $(0.6,\ 0)$, indicating that not all phase transitions lead to a divergent QFI—its divergence is governed by the closing of the energy gap.

\section{MEASUREMENT SCHEMES} \label{section:5}
\begin{widetext}
	
	\begin{figure}[h!]
		\centering\includegraphics[width=0.95\textwidth]{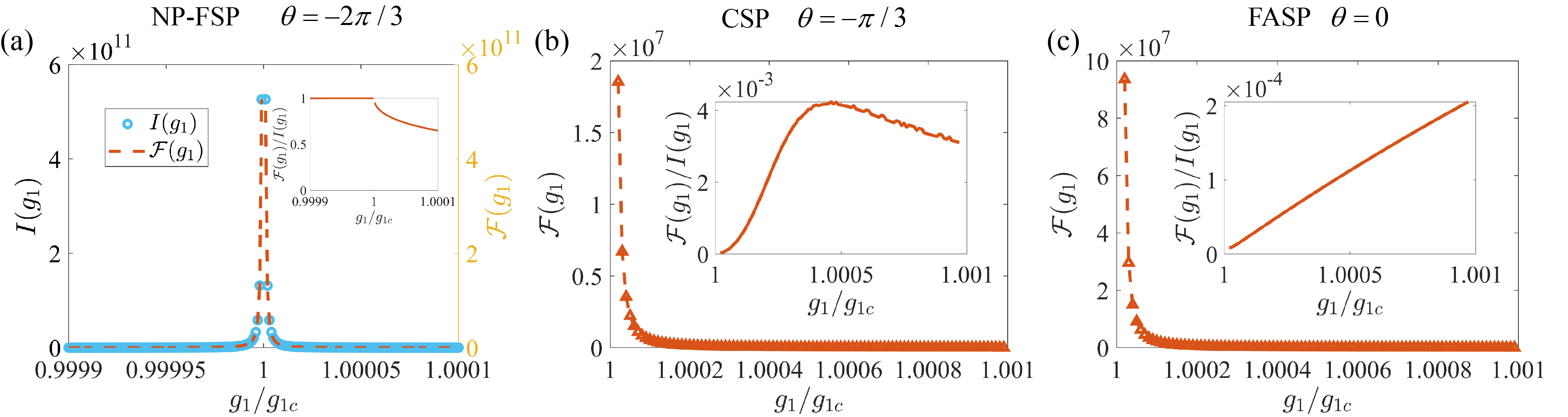}
		\caption{(a) The QFI $I(g_1)$ and the inverted variance $\mathcal{F}(g_1)$ as a function of $g_1$ normalized by $g_{1c}$ around the critical point $(g_{1c}(0,-2\pi/3),-2\pi/3)$ for the FSP-CSP transition. The blue dot line stands for $I(g_1)$ and the orange dashed line stands for $\mathcal{F}(g_1)$. The inset shows the ratio of ${\mathcal{F}}(g_1)$ to $I(g_1)$. (b) The inverted variance $\mathcal{F}(g_1)$ as a function of $g_1$ normalized by $g_{1c}$ around the critical point $(g_{1c}(-2\pi/3,-\pi/3),-\pi/3)$ in the CSP. The inset shows the ratio of ${\mathcal{F}}(g_1)$ to $I(g_1)$. (c) The inverted variance $\mathcal{F}(g_1)$ as a function of $g_1$ normalized by $g_{1c}$ around the critical point $(g_{1c}(-2\pi/3,0),0)$ in the FASP. The inset shows the ratio of ${\mathcal{F}}(g_1)$ to $I(g_1)$. } \label{Fig:fig6}
	\end{figure}
	
\end{widetext}

The QFI provides an upper bound for parameter estimation, the so-called quantum Cramér-Rao bound (QCRB) \cite{polino2020photonic}. In realistic physical implementations, an appropriate measurement operator must be selected to saturate this bound. According to the error propagation formula, the inverted variance for the estimation of  parameter $\lambda$ is given by \cite{wiseman_milburn_2009,di2023critical}
\begin{equation}\label{eq24}
	\mathcal{F}(\lambda)=\left(\frac{\partial_{\lambda}\langle\hat{O}\rangle}{\langle\Delta\hat{O}\rangle}\right)^2,
\end{equation}
where $\hat{O}$ is the chosen observable, and $\Delta\hat{O}=\hat{O}-\langle O\rangle$. In the QRT model, We choose the photon number operator of the first cavity $\hat{N}=\hat{N}_1=a_1^{\dagger}a_1$ as the measurement operator. In the NP, the average photon number in the ground state $\prod_{q}\ket{0}_q$ is given by,
\begin{eqnarray}\label{eq25}
	\langle\hat{N}_1\rangle&=&\frac{1}{3}(\nu_0^2+2\nu_{2\pi/3}^2),
\end{eqnarray}
and the variance of $\hat{N}$ is
\begin{eqnarray}\label{eq26}
	\langle\Delta\hat{N}\rangle^2&=&\langle\hat{N}^2\rangle-\langle\hat{N}\rangle^2\nonumber\\
	&=&\frac{2}{9}(\mu_0^2\nu_0^2+\mu_{2\pi/3}^2\nu_{2\pi/3}^2).
\end{eqnarray}
Inserting Eqs. (\ref{eq25}) and (\ref{eq26}) into Eq. (\ref{eq24}), we obtain the inverted variance of the parameter $g_1$,
\begin{eqnarray}
	\mathcal{F}(g_1)=\frac{(\partial_{g_1}(\nu_0^2+2\nu_{2\pi/3}^2))^2}{2(\mu_0^2\nu_0^2+\mu_{2\pi/3}^2\nu_{2\pi/3}^2)}.
\end{eqnarray}
Similarly, in the FSP, 
\begin{eqnarray}\label{eq27}
	\langle\hat{N}\rangle&=&(\tilde{\nu}_0^2+2\tilde{\nu}_{2\pi/3}^2)/3+|\alpha_1|^2,\nonumber\\
	\langle\Delta\hat{N}\rangle^2&=&\frac{2}{9}(\mu_0^2\nu_0^2+\mu_{2\pi/3}^2\nu_{2\pi/3}^2).
\end{eqnarray}
Inserting Eqs. (\ref{eq27}) into Eq. (\ref{eq24}), the inverted variance of the parameter $g_1$ is given by
\begin{eqnarray}
	\mathcal{F}(g_1)=\frac{(\partial_{g_1}(\tilde{\nu}_0^2+2\tilde{\nu}_{2\pi/3}^2)+3\partial_{g_1}|\alpha_1|^2)^2}{2(\tilde{\mu}_0^2\tilde{\nu}_0^2+\tilde{\mu}_{2\pi/3}^2\tilde{\nu}_{2\pi/3}^2)+6(\tilde{\nu}_0^2+2\tilde{\nu}_{2\pi/3}^2)+9|\alpha_1|^2}.\nonumber\\
\end{eqnarray}
These results are presented in Fig. \ref{Fig:fig6}(a) for $\theta=-2\pi/3$. As shown in the inset, $\mathcal{F}(g_1)$ consistently saturates the QCRB in the NP. In the FSP, it nearly reaches the QCRB near the critical point $g_{1c}(0,-2\pi/3)$, but the ratio of ${\mathcal{F}}(g_1)$ to $I(g_1)$ gradually decreases as the system goes away from the critical point.

In the CSP and FASP, the average photon number in the first cavity for the ground state $\prod_{i}\ket{0}_i$ is given by
\begin{eqnarray}
	\langle\hat{N}_1\rangle&=&\sum_{i=1}^{3}|T_{1,i}|^2+|\alpha_1|^2,
\end{eqnarray}
where $T_{n,i}$ denotes the matrix element of the transformation matrix $T$, and the expression of $\langle\hat{N}_1^2\rangle$ is presented in Appendix \ref{appendix5}. From Eq. (\ref{eq24}), ${\mathcal{F}}(g_1)$ can be calculated numerically. The plots of ${\mathcal{F}}(g_1)$ and the ratio of ${\mathcal{F}}(g_1)$ to $I(g_1)$ are shown in Figs. \ref{Fig:fig6}(b) and (c). It is observed that the inverted variance also diverges around the critical point $g_{1c}(-2\pi/3,-\pi/3)$ and $g_{1c}(-2\pi/3,0)$, but different from the NP and the FSP, it cannot saturate the QCRB.

\section{summary}\label{section:6}

In conclusion, we have investigated the criticality-enhanced quantum sensing of a QRT model. Our results reveal that the QFI exhibits pronounced peaks around the phase boundaries, indicating that quantum phase transitions provide a valuable resource for enhancing the precision of parameter estimation. By tuning the scaled coupling strength $g_1$ and the hopping phase $\theta$ controlled by an artificial magnetic field, we demonstrate that the QFI diverges as $|g_1-g_{1c}|^{-\nu}$ and $|\theta-(-\theta_c)|^{-\nu}$ in the phases near the second-order phase transition point and the triple point, with the divergence characterized by distinct critical exponent $\nu$. In contrast, the QFI is enhanced but remains finite near the first-order phase transition point. These results reveal that the divergence of the QFI is intrinsically associated with the energy gap closing. We further consider the resources consumed during the estimation process and find out that the divergent QFI consistently exhibits HL $\sim\langle N\rangle^2T^2$ for $\theta\neq0$, highlighting the significant contribution of the artificial magnetic field to the enhanced precision. By the use of the average photon number measurement scheme, we show that the QFI always saturates the QCRB in the NP. Our results demonstrate the enhancement of precision enabled by the artificial magnetic field and further reveal the considerable potential of few-body systems for critical quantum metrology.

\begin{acknowledgments}
	We thank Yuyu Zhang from CQU for helpful comments. This research is supported by the National Natural Science Foundation of China (NSFC) (Grants No. 12204371, 12474363), Shaanxi Fundamental Science Research Project for Mathematics and Physics (Grants No. 23JSQ012), Fundamental Research Funds for the Central Universities (Grant No. xxj032025044).	\\
\end{acknowledgments}

	\appendix
	
		\section{SW transformation of the original Hamiltonian}\label{appendix1}
	
	We consider the general expression of the original Hamiltonian (\ref{eq1}) with the displacement transformation (\ref{eq7}), which can be divided into two parts\cite{PhysRevLett.127.063602}
	\begin{equation}\label{A1}
		H_{\mathrm{QRT}}=H_s+H_1,
	\end{equation}  
	with the atom part 
	\begin{equation}
		H_s=\sum_{n=1}^3g\left(\alpha_n^*+\alpha_n\right)\sigma_n^x+\frac{\Delta}{2}\sigma_n^z,
	\end{equation}
	and the remaining part
	\begin{eqnarray}
		H_{1}&=&\sum_{n=1}^3\omega(\tilde{a}_n^\dagger+\alpha_n^*)(\tilde{a}_n+\alpha_n)+\sum_{n=1}^Ng\left(\tilde{a}_n^\dagger+\tilde{a}_n\right)\sigma_n^x\nonumber\\
		&\ &+J\sum_{n=1}^N(\tilde{a}_n^\dagger+\alpha_n^*)\nonumber\\
		&\ &\times[e^{i\theta}(\tilde{a}_{n+1}+\alpha_{n+1})+e^{-i\theta}(\tilde{a}_{n-1}+\alpha_{n-1})].\nonumber\\
	\end{eqnarray}
	
	To diagonalize $H_s$, we apply the transformation to the Pauli matrices: $\tau_z^n=\Delta/\Delta_n\sigma_n^z+4gA_n/\Delta_n\sigma_n^x$. Then the Hamiltonian (\ref{eq1}) becomes
	\begin{equation}
		H_{\mathrm{QRT}}=\sum_{n=1}^3H_{R,n}+J\tilde{a}_{n}^{\dagger}(e^{i\theta}\tilde{a}_{n+1}+e^{-i\theta}\tilde{a}_{n-1})+E_{0},
	\end{equation}
	where $E_0=\sum_{n=1}^{3}\omega\alpha_{n}^{*}\alpha_{n}+J\alpha_{n}^{*}(e^{i\theta}\alpha_{n+1}+e^{-i\theta}\alpha_{n-1})$ and the transformed quantum Rabi Hamiltonian is 
	\begin{equation}\label{A5}
		H_{R,n}=\omega\tilde{a}_{n}^{\dagger}\tilde{a}_{n}+\frac{\Delta_{n}}{2}\tau_{n}^{z}+\lambda_{n}\left(\tilde{a}_{n}^{\dagger}+\tilde{a}_n\right)\tau_{n}^{x},
	\end{equation}
	where the effective coupling strength is $\lambda_n=g\Delta/\Delta_n$.
	
	Then, we perform a Schrieffer–Wolff (SW) transformation $S_n=\text{exp}[-i\sigma_n^{y}\lambda_n/\Delta_n(\tilde{a}_n^{\dagger}+\tilde{a}_n)]$ on the Hamiltonian (\ref{A5}) \cite{hwang2015quantum}
	\begin{eqnarray}\label{A6}
		H^{\prime}_{R,n}&=&S^{\dagger}_nH_{R,n}S_n\nonumber\\
		&=&\omega \tilde{a}_n^{\dagger}\tilde{a}_n+\frac{\Delta_n}{2}\tau_n^{z}+\frac{\lambda_n^2}{\Delta_n}(\tilde{a}_n+\tilde{a}_n^{\dagger})^2\tau_n^z+O(\frac{\lambda_n^4}{\Delta_n^4}).\nonumber\\
	\end{eqnarray}
	
	Using the unitary transformation $U = \prod_{n=1}^{3} S_n$, we obtain the effective quantum Rabi triangle Hamiltonian
	\begin{equation}
		H_{\mathrm{QRT}}=\sum_{n=1}^3H_{R,n}'+J\sum_{n=1}^3(e^{i\theta}\tilde{a}_n^\dagger \tilde{a}_{n'}+h.c.)+E_0.
	\end{equation}
	In the $\Delta/\omega\rightarrow\infty$ limit, the higher-order terms of Eq. (\ref{A6}) can be neglected. Therefore, the lower-energy Hamiltonian after projection is
	\begin{eqnarray}
		H_{\mathrm{eff}}&=&\sum_{n=1}^3\omega\tilde{a}_n^\dagger\tilde{a}_n-\frac{\lambda_n^2}{\Delta_n}\left(\tilde{a}_n^{\dagger}+\tilde{a}_n\right)^2\nonumber\\
		&\ &+J\tilde{a}_n^\dagger(e^{i\theta}\tilde{a}_{n+1}+e^{-i\theta}\tilde{a}_{n-1})+E_0.
	\end{eqnarray}
	
	In the NP, $\alpha_n=0$, therefore the effective Hamiltonian can be simplified as
	\begin{eqnarray}
		H_{\mathrm{NP}}&=&\sum_{n=1}^{3}\omega a_{n}^{\dagger}a_{n}-g_{1}^{2}\omega\left(a_{n}^{\dagger}+a_{n}\right)^{2}\nonumber\\
		&\ &+Ja_{n}^{\dagger}(e^{i\theta}a_{n+1}+e^{-i\theta}a_{n-1})-\frac{3\Delta}{2}.
	\end{eqnarray}

	\begin{widetext}		
		
	\section{Diagonalization of the Hamiltonian}\label{appendix2}
	
	  In the NP, we apply the Bogoliubov transformation 
	\begin{equation}
		a_q=\mu_qb_q+\nu_qb_{-q}^{\dagger},\ \ \ a_{-q}=\mu_{q}b_{-q}+\nu_{q}b_q^{\dagger},
	\end{equation}
	since the operator $a_q$ and $b_q$ must satisfy the bosonic commutation relations, we set $\mu_q=\cosh(\xi_q)$ and $\nu_q=\sinh(\xi_q)$. Then the transformed Hamiltonian is 	
	\begin{eqnarray}
		H_{\mathrm{NP}}&=&\sum_q \left(-2\omega g_1^2 \cosh(\xi_q) \sinh(\xi_{q}) + \omega_q\cosh^2(\xi_q)  \right)b_q^\dagger b_q +  \left(-2\omega g_1^2 \cosh(\xi_{q}) \sinh(\xi_q) +  \omega_q\sinh^2(\xi_q) \right)b_{-q}^\dagger b_{-q}\nonumber\\
		&\ &+ \frac{1}{2}\left(-2\omega g_1^2 ( \cosh^2(\xi_q)+ \sinh^2(\xi_q))+ (\omega_q+\omega_{-q})\cosh(\xi_q) \sinh(\xi_q)  \right)(b_{-q} b_q+ b_{-q}^\dagger b_q^\dagger)\nonumber\\
		&\ &+\frac{(\omega_q+\omega_{-q})}{2}\sinh^2(\xi_q) -2g_1^2\omega\cosh(\xi_q) \sinh(\xi_{q}).
	\end{eqnarray}    
	To eliminate the off-diagonal terms, it requires that 
	\begin{equation}
		\xi_q=\frac{1}{4}\ln\frac{\omega_q+\omega_{-q}+4\omega g_1^2}{\omega_q+\omega_{-q}-4\omega g_1^2},
	\end{equation}
	then the diagonalized Hamiltonian is
	\begin{equation}
		H_{\mathrm{NP}}=\sum_q\epsilon_q b_q^{\dagger}b_q+E_q,
	\end{equation}
	with
	\begin{eqnarray}
		\epsilon_q&=&\frac{1}{2}(\sqrt{(\omega_q+\omega_{-q})^2-16\omega^2g_1^4}+\omega_q-\omega_{-q}),\nonumber\\
		E_q&=&\frac{1}{4} ( \sqrt{(\omega_q+\omega_{-q})^2-16\omega^2g_1^4}  -(\omega_{q} + \omega_{-q})).
	\end{eqnarray}
	In this mode, the ground state is $\prod_{q}\ket{0}_{q}$ with the ground state energy $E_g=\sum_qE_q$. The eigenstates is $\prod_{q}\ket{n_q}_{q}=\prod_{q}(b^{\dagger}_q)^{ n_q}\ket{0}_{q}$ with corresponding eigenvalues $E_n=E_g+\sum_qn_q\epsilon_q$.
		
	In the SP, according to the huge increase of photon populations, we first apply the displacement (\ref{eq7}) to the Hamiltonian (\ref{eq1}). Considering the Hamiltonian (\ref{A1}), eliminating the off-diagonal terms requires that\cite{PhysRevLett.127.063602}
	\begin{eqnarray}\label{B7}
		\omega A_n-\frac{4g^2A_n}{\sqrt{16g^2A_n^2+\Delta^2}}+J\cos\theta(A_{n+1}+A_{n-1})+J\sin\theta(B_{n-1}-B_{n+1})&=&0,\nonumber\\
		\omega B_n+J\sin\theta(A_{n+1}-A_{n-1})+J\cos\theta(B_{n+1}+B_{n-1})&=&0.
	\end{eqnarray}
	We can obtain that $\sum_{n}^{3}B_n=0$. Assume $B_1=0$, then $B_2=-B_3$,
	\begin{eqnarray}
		\omega B_1+J\sin\theta(A_{2}-A_{0})+J\cos\theta(B_{2}+B_{3})&=&0,\nonumber\\
		\omega B_2+J\sin\theta(A_{3}-A_{1})+J\cos\theta(B_{3}+B_{1})&=&0.
	\end{eqnarray}
	Assuming $A_2=A_3=a,\ A_1=b$, we derive the relations of $A_n$,
	\begin{eqnarray}\label{B8}
		(\omega -\frac{4g^2}{\sqrt{16g^2b^2+\Delta^2}}-\frac{2J^2\sin^2\theta}{\omega-J\cos\theta})b+2(J\cos\theta+\frac{J^2\sin^2\theta}{\omega-J\cos\theta})a&=&0,\nonumber\\
		(\omega -\frac{4g^2}{\sqrt{16g^2a^2+\Delta^2}}-\frac{2J^2\sin^2\theta}{\omega-J\cos\theta})a+(J\cos\theta+\frac{J^2\sin^2\theta}{\omega-J\cos\theta})(a+b)&=&0.
	\end{eqnarray}
	
	In the FSP, the displacement $\alpha_n$ for each cavity is real and identical. Therefore, the solution of Eqs. (\ref{B8}) is
	\begin{equation}
		a=b=A_n=\pm\sqrt{\frac{g^2}{(\omega+2J\cos\theta)^2}-\frac{\Delta^2}{16g^2}}
	\end{equation} 
	We consider the Hamiltonian (\ref{eq11}) and similarly apply the Bogoliubov transformation:
	\begin{equation}
		\tilde{a}_q=\tilde{\mu}_q\tilde{b}_q+\tilde{\nu}_q\tilde{b}_{-q}^{\dagger},\ \ \ \tilde{a}_{-q}=\tilde{\mu}_{q}\tilde{b}_{-q}+\tilde{\nu}_{q}\tilde{b}_q^{\dagger},
	\end{equation}
	with the operator $\tilde{\mu}_q=\cosh({\xi}^{\prime}_q)$ and $\tilde{\nu}_q=\sinh({\xi}^{\prime}_q)$ satisfying the condition 
	\begin{equation}
		{\xi}^{\prime}_q=\frac{1}{4}\ln\frac{\omega_q^{\prime}+\omega_{-q}^{\prime}+4{\lambda^{'2}}/{\Delta^{\prime}}}{\omega_q^{\prime}+\omega_{-q}^{\prime}-4{\lambda^{'2}}/{\Delta^{\prime}}},
	\end{equation}
	then the diagonalized Hamiltonian takes the form  
	\begin{equation}
		H_{\mathrm{FSP}}=\sum_q\epsilon_q^{\prime} \tilde{b}_q^{\dagger}\tilde{b}_q+E_q^{\prime},
	\end{equation}
	with
	\begin{eqnarray}
		\epsilon_q^{\prime}&=&\frac{1}{2}(\sqrt{(\omega_q^{\prime}+\omega^{\prime}_{-q})^2-16{\lambda^{'4}}/{\Delta^{\prime2}}}+\omega^{\prime}_q-\omega^{\prime}_{-q})\nonumber\\
		E_{q}^{\prime}&=&-\frac{\lambda^{'2}}{\Delta^{\prime}}+\frac{1}{2}\sum_{q}(\epsilon_{q}^{\prime}-\omega_{q}^{\prime}).
	\end{eqnarray}
	In this mode, the ground state is $\prod_{q}\ket{0}_{q}$ with the ground state energy $E_g^{\prime}=\sum_qE_q^{\prime}$. The eigenstates is $\prod_{q}\ket{n_q}_{q}=\prod_{q}(\tilde{b}^{\dagger}_q)^{ n_q}\ket{0}_{q}$ with corresponding eigenvalues $E_n^{\prime}=E_g^{\prime}+\sum_qn_q\epsilon_q^{\prime}$.
	
	In the CSP, since $\Delta_n\neq\Delta_{n^{\prime}}$, the Hamiltonian cannot be transformed into $q$ space. The solutions of Eqs. (\ref{B7}) correspond to the minimum of the ground-state energy and are obtained numerically. The ground-state energy is expressed by
	\begin{equation}
		E_g=\sum_{n=1}^N\omega(A_n^2+B_n^2)-\frac12\sqrt{\Delta^2+16g^2A_n^2}+2J[(A_nA_{n+1}+B_nB_{n+1})\cos\theta+\sin\theta(B_nA_{n+1}-B_{n+1}A_n)].
	\end{equation}
	Then we diagonalize the Hamiltonian (\ref{eq10}) using the Bogoliubov transformation. With the notation $\alpha=\{\tilde{a}_1,\tilde{a}_2,\tilde{a}_3, \tilde{a}_1^{\dagger},\tilde{a}_2^{\dagger},\tilde{a}_3^{\dagger}\}$, the Hamiltonian reads 
	\begin{equation}\label{B13}
		H_{\mathrm{CSP}}=\alpha M\alpha^{\dagger} ,
	\end{equation}
	where 			
	\begin{eqnarray}
		M = \begin{pmatrix}
			\omega/2 - \lambda_1^2/\Delta_1 & Je^{-i \theta}/2 & Je^{i \theta}/2 & -\lambda_1^2/\Delta_1 & 0 & 0 \\
			Je^{i \theta}/2 & \omega/2 - \lambda_2^2/\Delta_2 & Je^{-i \theta}/2 & 0 & -\lambda_2^2/\Delta_2 & 0 \\
			Je^{-i \theta}/2 & Je^{i \theta}/2 & \omega/2 - \lambda_3^2/\Delta_3 & 0 & 0 & -\lambda_3^2/\Delta_3 \\
			-\lambda_1^2/\Delta_1 & 0 & 0 & \omega/2 - \lambda_1^2/\Delta_1 & Je^{i \theta}/2 & Je^{-i \theta}/2 \\
			0 & -\lambda_2^2/\Delta_2 & 0 & Je^{-i \theta}/2 & \omega/2 - \lambda_2^2/\Delta_2 & Je^{i \theta}/2 \\
			0 & 0 & -\lambda_3^2/\Delta_3 & Je^{i \theta}/2 & Je^{-i \theta}/2 & \omega/2 - \lambda_3^2/\Delta_3
		\end{pmatrix}.
	\end{eqnarray}
	We set $\alpha^{\dagger}=T \beta^{\dagger}$, where $\beta=\{\tilde{c}_1^{\dagger},\tilde{c}_2^{\dagger},\tilde{c}_3^{\dagger}, \tilde{c}_1,\tilde{c}_2,\tilde{c}_3\}$ and T is a $6\times6$ matrix, $\tilde{a}_n^{\dagger}=\sum_{i=1}^3T_{n,i}\tilde{c}_i+T_{n,i+3}\tilde{c}_i^{\dagger}$
	, $H_{\mathrm{CSP}}=\beta T^{\dagger} M T\beta^{\dagger}=2\sum_n^3\epsilon_n\tilde{c}^{\dagger}_n\tilde{c}_n+(\epsilon_n-\omega)/2$. For bosonic operators, it requires that $T\Lambda_-T^{\dagger}=\Lambda_-$, with $\Lambda_-=\mqty(I_{3\times3}&\ &0\\ 0&\ &-I_{3\times3})$. Then $\Lambda_-T^{\dagger}MT=\Lambda_-T^{\dagger}\Lambda_-\Lambda_-MT=T^{-1}\Lambda_-MT$, therefore, the eigenvalues $\epsilon_n$ can be obtained by diagonalizing $\Lambda_-M$. In this mode, the ground state is $\prod_{i}\ket{0}_{i}$ and the eigenstates is $\prod_{i}\ket{n_i}_{i}=\prod_{i}(\tilde{c}^{\dagger}_i)^{ n_i}\ket{0}_{i}$.
	
	 \section{The quantum Fisher information of the parameter $g_1$}\label{appendix3}
	 
	 In the normal phase, the partial derivative of Hamiltonian (\ref{eq6}) with respect to $g_1$ is
	 \begin{eqnarray}\label{C1}
	 	\partial_{g_1}H_{\mathrm{NP}}&=&\sum_q-2g_1\omega(2a_q^{\dagger}a_q+a_qa_{-q}+a^{\dagger}_qa^{\dagger}_{-q})\nonumber\\
	 	&=&\sum_q-2g_1\omega(\mu_q+\nu_q)^2[(b_qb_{-q}+b_q^{\dagger}b_{-q}^{\dagger})+2b_q^{\dagger}b_{q}+1].
	 \end{eqnarray}
	 
	 Substituting Eq. (\ref{C1}) into Eq. (\ref{eq17}) yields the expression of the QFI 
	 \begin{eqnarray}\label{C2}
	 	I_{\mathrm{NP}}(g_1)&=&4\sum_n\frac{|\prod_{q}\bra{0}_{q}\sum_q-2g_1\omega[(\mu_q+\nu_q)^2(b_qb_{-q}+b_q^{\dagger}b_{-q}^{\dagger})+2(\mu_q+\nu_q)^2b_q^{\dagger}b_{q}+2\nu_q^2+2\mu_q\nu_q]\prod_{q}(b^{\dagger}_q)^{ n_q}\ket{0}_{q}|^2}{(E_n-E_g)^2}\nonumber\\
	 	&=&4\sum_{q}\frac{|\prod_{q}\bra{0}_{q}-2g_1\omega(\mu_q+\nu_q)^2b_qb_{-q}b_q^{\dagger}b_{-q}^{\dagger}\ket{0}_q|^2}{(E_{1_q,1_{-q}}-E_g)^2}\nonumber\\
	 	&=&{16\omega^2g_1^2}\sum_q\frac{2}{(\omega_q+\omega_{-q}-4\omega g_1^2)^2},
	 \end{eqnarray}
	 where $E_{1_q,1_{-q}}=E_g+\epsilon_q+\epsilon_{-q}$.
	 
	 In the FSP,  the partial derivative of Hamiltonian (\ref{eq11}) with respect to $g_1$ is
	 \begin{eqnarray}
	 	\partial_{g_1}H_{\mathrm{FSP}} &=& \sum_q-\partial_{g_1}(\frac{\lambda^{\prime 2}}{\Delta^{\prime}})(2\tilde{a}_q^\dagger \tilde{a}_q+(\tilde{a}_q\tilde{a}_{-q}+\tilde{a}_q^\dagger \tilde{a}_{-q}^\dagger))+\partial_{g_1}(E_0-\frac{\Delta^{\prime}}{2})\nonumber\\
	 	&=&\sum_q-\partial_{g_1}(\frac{\lambda^{\prime 2}}{\Delta^{\prime}})[(\mu_q+\nu_q)^2(\tilde{b}_q\tilde{b}_{-q}+\tilde{b}_q^{\dagger}\tilde{b}_{-q}^{\dagger})+2(\mu_q+\nu_q)^2\tilde{b}_q^{\dagger}\tilde{b}_{q}]-\partial_{g_1}(\frac{\lambda^{\prime 2}}{\Delta^{\prime}})(\mu_q+\nu_q)^2+\partial_{g_1}(E_0-\frac{\Delta^{\prime}}{2}).\nonumber\\
	 \end{eqnarray}
	 The QFI
	 \begin{eqnarray}
	 	I_{\mathrm{FSP}}(g_1)
	 	&=&\frac{(2 J \cos \theta +\omega )^6}{64 g_1^{10} \omega ^4}\sum_q\frac{2}{(\omega_q^{\prime}+\omega_{-q}^{\prime}-4{\lambda^{'2}}/{\Delta^{\prime}})^2}.
	 \end{eqnarray}
	 In the CSP,  the partial derivative of Hamiltonian (\ref{eq10}) with respect to $g_1$ is 
	 \begin{eqnarray}
	 	\partial_{g_1}H_{\mathrm{CSP}}&=&\sum_{n=1}^3\partial_{g_1}(\frac{\lambda^{2}_n}{\Delta_{n}})(\tilde{a}^{\dagger}_n+\tilde{a}_n)^2=\sum_{n=1}^3\partial_{g_1}(\frac{\lambda^{2}_n}{\Delta_{n}})(\sum_{i=1}^{3}((T_{n,i}+T_{n,i+3}^*)\tilde{c}_i+(T_{n,i+3}+T_{n,i}^*)\tilde{c}_i^{\dagger}))^2\nonumber\\
	 	&=&\sum_{n=1}^3\partial_{g_1}(\frac{\lambda^{2}_n}{\Delta_{n}})\sum_{i=1}^{3} \sum_{j=1}^{3}[(T_{n,i} + T_{n,i+3}^*)(T_{n,j} + T_{n,j+3}^*) \tilde{c}_i \tilde{c}_j + (T_{n,i} + T_{n,i+3}^*)(T_{n,j+3} + T_{n,j}^*) \tilde{c}_i \tilde{c}_j^{\dagger} \nonumber\\
	 	&\ &+ (T_{n,i+3} + T_{n,i}^*)(T_{n,j} + T_{n,j+3}^*) \tilde{c}_i^{\dagger} \tilde{c}_j+ (T_{n,i+3} + T_{n,i}^*)(T_{n,j+3} + T_{n,j}^*) \tilde{c}_i^{\dagger} \tilde{c}_j^{\dagger}].
	 \end{eqnarray}
	Then the QFI
	 \begin{eqnarray}
	 	I_{\mathrm{CSP}}(g_1)
	 	&=&4\sum_{i=1}^{3}(\frac{2(\sum_{n=1}^3\partial_{g_1}(\frac{\lambda^{2}_n}{\Delta_{n}})(T_{n,i} + T_{n,i+3}^*)^2)(\sum_{m=1}^3\partial_{g_1}(\frac{\lambda^{2}_m}{\Delta_{m}})(T_{m,i+3} + T_{m,i}^*)^2)}{4\epsilon_i^2}\nonumber\\
	 	&\ &+4\sum_{j=1(j\neq i)}^3\frac{(2\sum_{n=1}^3\partial_{g_1}(\frac{\lambda^{2}_n}{\Delta_{n}})(T_{n,i} + T_{n,i+3}^*)(T_{n,j} + T_{n,j+3}^*))(2\sum_{m=1}^3\partial_{g_1}(\frac{\lambda^{2}_m}{\Delta_{m}})(T_{m,i+3} + T_{m,i}^*)(T_{m,j+3} + T_{m,j}^*))}{(\epsilon_i+\epsilon_j)^2}).\nonumber\\
	 \end{eqnarray}

	 \section{Approximate adiabatic evolution in the normal phase}\label{appendix4}
	
	 In the normal phase, to investigate the adiabatic preparation for the initial ground state, we consider a time-dependent Hamiltonian with a slowly varying scaled coupling strength $g_1$ \cite{PhysRevLett.124.120504},
	\begin{equation}
		H_{\mathrm{NP}}(t)=\sum_q\omega_q(t)a_q^\dagger a_q-g_1(t)^2\omega(a_qa_{-q}+a_q^\dagger a_{-q}^\dagger).
	\end{equation}
	Its instantaneous eigenstates are given by the squeezed Fock state
	\begin{eqnarray}
		H_{\mathrm{NP}}(t)\ket{\psi_n(t)}=\epsilon_q(t)\ket{\psi_n(t)},\ \ \ \ \ket{\psi_n(t)}=\sum_qe^{\xi_q(t)(a_q^{\dagger}a_{-q}^{\dagger}-a_{q}a_{-q})/2}\ket{n},
	\end{eqnarray}
	where 
	\begin{equation}
		\xi_q(t)=\frac{1}{4}\ln\frac{2\omega+2J\cos(\theta-q)+2J\cos(\theta+q)}{2\omega+2J\cos(\theta-q)+2J\cos(\theta+q)-8\omega g_1(t)^2},
	\end{equation}
	with the energy gap 
		\begin{eqnarray}
			\epsilon_q(t)&=&\frac{1}{2}[2J\cos(\theta-q)-2J\cos(\theta+q)\nonumber\\
			&\ &+\sqrt{(2\omega+2J\cos(\theta-q)+2J\cos(\theta+q))^2-4\omega g_1(t)^2(2\omega+2J\cos(\theta-q)+2J\cos(\theta+q))}].
		\end{eqnarray}
	\end{widetext}
	
	The system state can be decomposed over the basis as
	\begin{equation}
		\ket{\psi(t)}=\sum_{n=0}^{\infty}\alpha_n(t)e^{-i\Theta_n(t)}\ket{n_s(t)},
	\end{equation}
	where 
	\begin{equation}
		\Theta_n(t)=\int_0^tn\epsilon_q(t^{\prime})dt^{\prime}.
	\end{equation}
	To keep the system in the ground state, the condition $\alpha_n = 0$ for $n \neq 0$ must be satisfied. The evolution of $\alpha_n$ is then governed by the Schrödinger equation,
	\begin{equation}\label{D7}
		\frac{d\alpha_n(t)}{dt}=-\sum_{m=0}^{\infty}\alpha_m(t)Ae^{-i\left[\Theta _{m}(t)-\Theta_{n}(t)\right]}\langle n_{s}(t)|\frac{\partial}{\partial t}|m _{s}(t)\rangle, 
	\end{equation}
	by \(\delta g_1 = v\delta t\), Eq. (\ref{D7}) is rewritten as	
	\begin{eqnarray}
		\alpha_n(g_1) &=& -\sum_{m=0}^{\infty} \int_0^{g_1} \alpha_m(g_1') e^{-i[\Theta_m(g_1') - \Theta_n(g_1')]}\nonumber\\
		&\ &\times \langle n_s(g_1')| \frac{\partial}{\partial g_1'}| m_s(g_1') \rangle dg_1'.
	\end{eqnarray}
	With the assumption $\alpha_m(0)=\delta_{1,m}$, 	
	\begin{equation}
		\alpha_n(g_1) = -\int_0^{g_1} e^{-i[\Theta_0(g_1') - \Theta_n(g_1')]} \langle n_s(g_1')| \frac{\partial}{\partial g_1'}| 0_s(g_1') \rangle dg_1', 
	\end{equation}
	with	
	\begin{eqnarray}
		&&\langle n_s(g_1')| \frac{\partial}{\partial g_1'}| 0_s(g_1') \rangle = \langle n| S^\dagger(g_1') \frac{\partial}{\partial g_1'} S(g_1') |0 \rangle\nonumber\\
		& &=\frac{\omega g_1'}{4(\omega+J\cos(\theta-q)+J\cos(\theta+q)-4\omega g_1'^2)} \delta_{n,2}.\nonumber\\
	\end{eqnarray}
	This indicates that only transitions to the second-excited state \(S(g_1')|2\rangle\) need to be taken into account,	
	\begin{equation}
		\alpha_2(t) = -\frac{1}{4} \int_0^g f(g_1') e^{iR(g_1')} dg_1',
	\end{equation}
	where we define 
	\begin{eqnarray}
		f(g_1)&=&\frac{\omega g_1}{\omega+J\cos(\theta-q)+J\cos(\theta+q)-4\omega g_1^2},\nonumber\\
		R(g_1)&=&\Theta_2(g_1)-\Theta_0(g_1)\nonumber\\
		=\int_0^{g_1}&dg_1'&\frac{2\sqrt{(2J\cos\theta\cos q+\omega )(-4 g_1'^2\omega +2J\cos\theta\cos q+\omega)}}{v(g_1')}.\nonumber\\
	\end{eqnarray}
	To maintain the adiabatic evolution process, $\alpha_2$ should remain small during the evolution. Therefore, we need $v(g)\ll1$ so that $R(g)$ is large and the exponential in the integral oscillates fast, cancelling the integral. This suggest the condition $\frac{\dot{f}}{f}\ll\dot{R}$, which leads to
	\begin{eqnarray}
		v(g_1')&\ll&2g_1'(2 J \cos\theta\cos q+\omega )^{1/2}( \omega+2 J \cos\theta \cos q +4 g_1'^2 \omega)^{-1}\nonumber\\
		&\ &\times ( \omega+2 J \cos\theta \cos q -4 g_1'^2 \omega)^{3/2},
	\end{eqnarray}
	when $g_1$ is approached $g_{1c}$, 
	\begin{equation}
		v(g_1')=\gamma( \omega+2 J \cos\theta \cos q -4 g_1'^2 \omega)^{3/2}\sim\gamma(g_{1c}-g_1)^{3/2},
	\end{equation}
	with $\gamma$ a small constant. 
	
	The adiabatic preparation time for the initial ground state is calculated as
	\begin{equation}
		T=\int_0^{g_1}\frac{1}{v(g_1')}dg_1'\sim\frac{1}{\gamma(g_{1c}-g_1)^{1/2}}\sim\frac{1}{\gamma\epsilon_0(g_1)}.
	\end{equation}
	
	\begin{widetext}	
		
	\section{The scaling of the average photon number and the first excitation energy}\label{appendix5}	
		
	\begin{figure}[h!]
		\centering\includegraphics[width=0.95\textwidth]{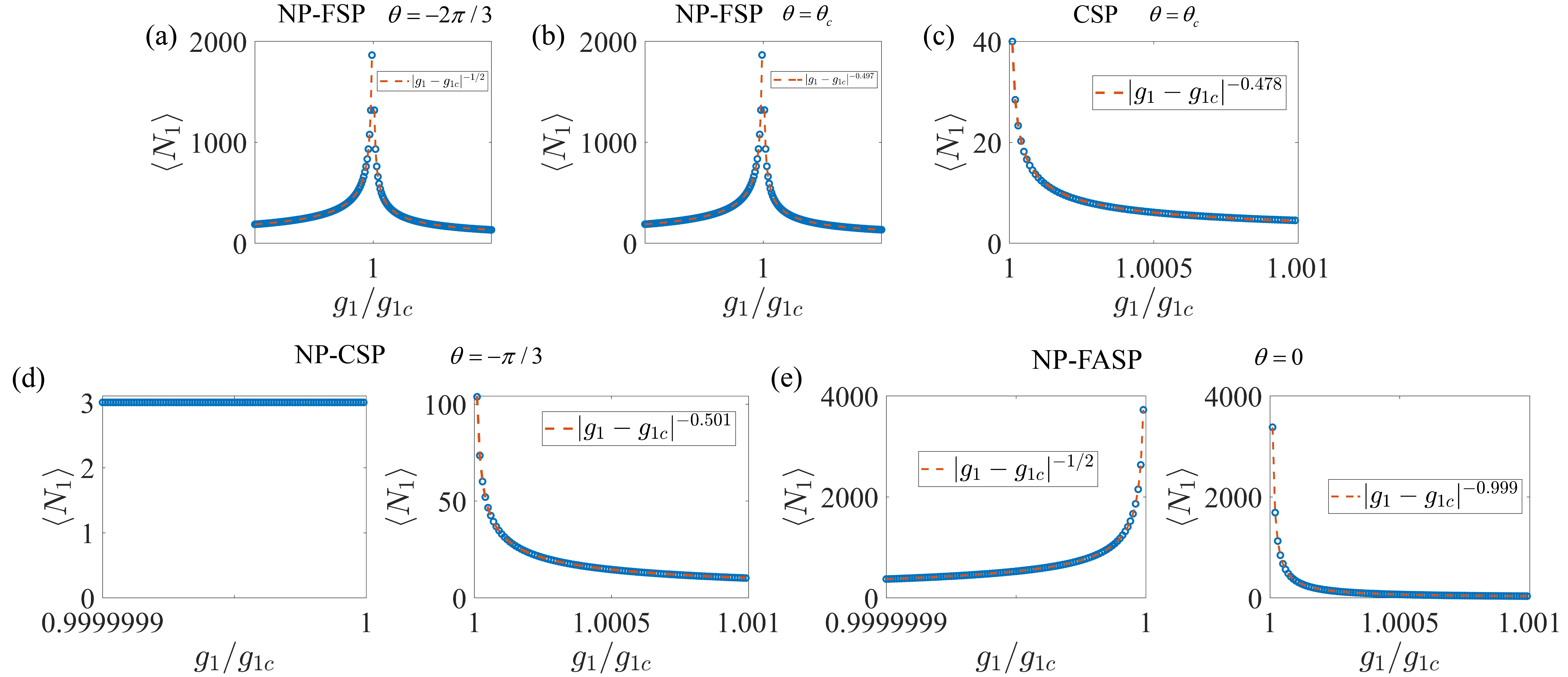}
		\caption{The scaling of the average photon number $\hat{N}_1$ as a function of $g_1$ normalized by $g_{1c}$ near the critical points. (a) In the NP-FSP transition with $\theta=-2\pi/3$. (b) In the NP-FSP transition with $\theta=-\theta_c$. (c) In the CSP with $\theta=-\theta_c$. (d) In the NP-CSP transition with $\theta=-\pi/3$. (e) In the NP-FASP transition with $\theta=0$.}\label{Fig:fig7}
	\end{figure}
	
	\begin{figure}[h!]
		\centering\includegraphics[width=0.95\textwidth]{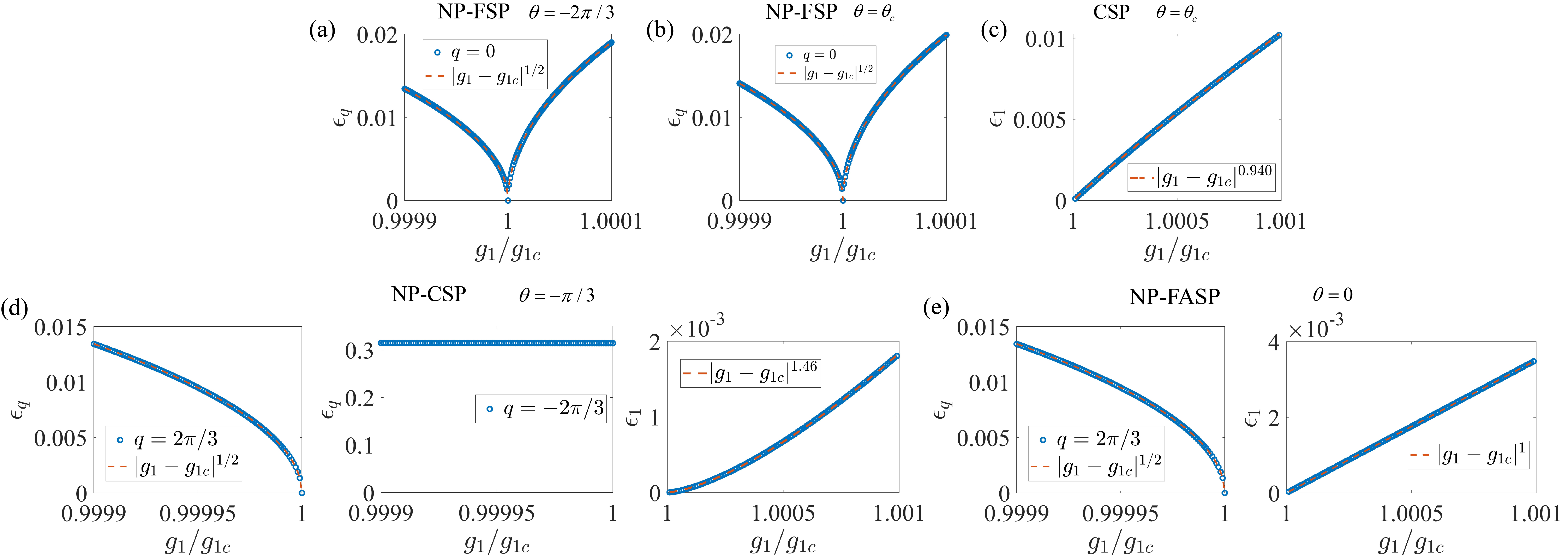}
		\caption{The scaling of the first excitation energy as a function of $g_1$ normalized by $g_{1c}$ near the critical points. (a) In the NP-FSP transition with $\theta=-2\pi/3$. (b) In the NP-FSP transition with $\theta=-\theta_c$. (c) In the CSP with $\theta=-\theta_c$. (d) In the NP-CSP transition with $\theta=-\pi/3$. (e) In the NP-FASP transition with $\theta=0$.}\label{Fig:fig8}
	\end{figure}
	
	\begin{figure}[h!]
		\centering\includegraphics[width=0.95\textwidth]{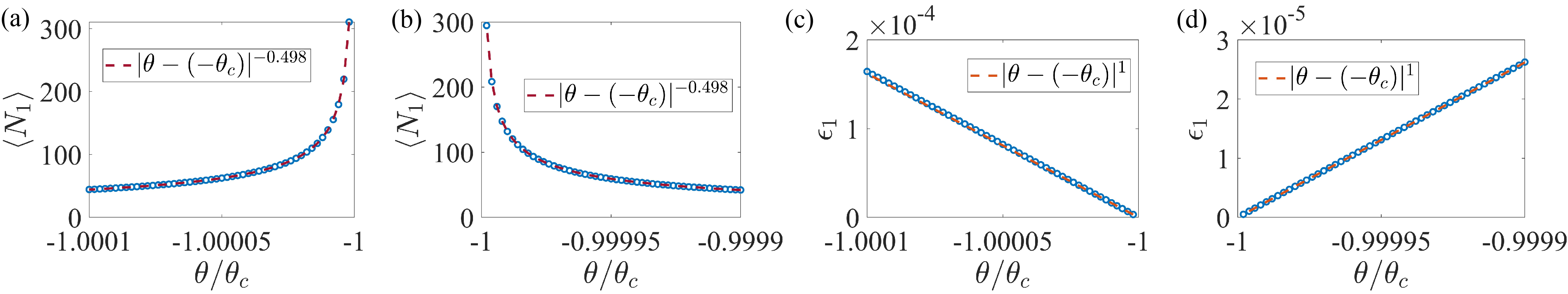}
		\caption{(a)(b) The scaling of the average photon number $\hat{N}_1$ as a function of $\theta$ normalized by $\theta_c$ near the critical point $(g_{1c}(0,-\theta_c),-\theta_c)$ of the FSP-CSP transition. (c)(d) The scaling of the first excitation energy as a function of $\theta$ normalized by $\theta_c$ around the critical point $(g_{1c}(0,-\theta_c),-\theta_c)$ of the FSP-CSP transition.}\label{Fig:fig9}
	\end{figure}
		
	 In the NP, the average photon number of the first cavity 
	 
	\begin{eqnarray}
		\hat{N}_1&=&a_1^{\dagger}a_1=\frac{1}{3}\sum_qa_q^{\dagger}a_q=\frac{1}{3}\sum_q\mu_q^2b_q^{\dagger}b_q+\nu_q^2b_{-q}b_{-q}^{\dagger}+\mu_q\nu_q(b_q^{\dagger}b_{-q}^{\dagger}+b_{-q}b_q),
	\end{eqnarray} 
	In the ground state $\prod_{q}\ket{0}_q$
	\begin{eqnarray}
		\langle\hat{N}_1\rangle&=&\frac{1}{3}(\nu_0^2+2\nu_{2\pi/3}^2).
	\end{eqnarray}
	In the FSP, 
	\begin{eqnarray}
		\hat{N}_1=(\tilde{a}_1^{\dagger}+\alpha_1^*)(\tilde{a}_1+\alpha_1)=\tilde{a}_1^{\dagger}\tilde{a}_1+\alpha_1\tilde{a}_1^{\dagger}+\alpha_1^*\tilde{a}_1+|\alpha_1|^2,
	\end{eqnarray}
	then $\langle\hat{N}_1\rangle=(\tilde{\nu}_0^2+2\tilde{\nu}_{2\pi/3}^2)/3+|\alpha_1|^2$. 
	
	In the CSP,
	\begin{eqnarray}
		\hat{N}_1&=&(\sum_{i=1}^{3}T_{1,i}\tilde{c}_i+T_{1,i+3}\tilde{c}_i^{\dagger}+\alpha_1^*)(\sum_{j=1}^{3}T_{1,j}^*\tilde{c}_j^{\dagger}+T_{1,j+3}^*\tilde{c}_j+\alpha_1),
	\end{eqnarray}
	and in the ground state $\prod_{i}\ket{0}_i$
	\begin{eqnarray}
		\langle\hat{N}_1\rangle&=&\sum_{i=1}^{3}|T_{1,i}|^2+|\alpha_1|^2.
	\end{eqnarray}
	The variance of $\hat{N_1}$
	\begin{eqnarray}
		\langle\hat{N}_1^2\rangle&=&|\alpha_1|^4+|T_{1,1}|^4+3|\alpha_1|^2|T_{1,1}|^2+2|T_{1,1}|^2|T_{1,2}|^2+2|T_{1,1}|^2|T_{1,3}|^2+2|T_{1,1}|^2|T_{1,4}|^2 \nonumber\\
		&\ &+|T_{1,1}|^2|T_{1,5}|^2+|T_{1,1}|^2|T_{1,6}|^2+T_{1,1}T_{1,4}T_{1,2}^*T_{1,5}^*+T_{1,1}T_{1,4}T_{1,3}^*T_{1,6}^*+\alpha_1^2T_{1,1}T_{1,4} \nonumber\\
		&\ &+T_{1,2}T_{1,5}T_{1,1}^*T_{1,4}^*+T_{1,3}T_{1,6}T_{1,1}^*T_{1,4}^*+(\alpha_1^\dagger)^2T_{1,1}^*T_{1,4}^*+|T_{1,2}|^4+3|\alpha_1|^2|T_{1,2}|^2 \nonumber\\
		&\ &+2|T_{1,2}|^2|T_{1,3}|^2+|T_{1,2}|^2|T_{1,4}|^2+2|T_{1,2}|^2|T_{1,5}|^2+|T_{1,2}|^2|T_{1,6}|^2 \nonumber\\
		&\ &+T_{1,2}T_{1,5}T_{1,3}^*T_{1,6}^*+\alpha_1^2T_{1,2}T_{1,5}+T_{1,3}T_{1,6}T_{1,2}^*T_{1,5}^*+(\alpha_1^\dagger)^2T_{1,2}^*T_{1,5}^* \nonumber\\
		&\ &+|T_{1,3}|^4+3|\alpha_1|^2|T_{1,3}|^2+|T_{1,3}|^2|T_{1,4}|^2+|T_{1,3}|^2|T_{1,5}|^2+2|T_{1,3}|^2|T_{1,6}|^2 \nonumber\\
		&\ &+\alpha_1^2T_{1,3}T_{1,6}+(\alpha_1^\dagger)^2T_{1,3}^*T_{1,6}^*+|\alpha_1|^2|T_{1,4}|^2+|\alpha_1|^2|T_{1,5}|^2+|\alpha_1|^2|T_{1,6}|^2
	\end{eqnarray}
	
	The scaling of $\hat{N}_1$ and the first excitation energy as functions of $g_1$ normalized by $g_{1c}$ are shown in Fig. \ref{Fig:fig7} and Fig. \ref{Fig:fig8}. From Fig. \ref{Fig:fig7}(d), it can be seen that when $\theta = -\pi/3$ in the normal phase, the average photon number remains finite. As shown in Fig. \ref{Fig:fig8}(d), only the energy gap $\epsilon_{-2\pi/3}$ closes, while $\epsilon_{2\pi/3}$ stays approximately constant. Consequently, the denominator of $I_{\mathrm{NP}}(g_1)$ in Eq. (\ref{C2}), namely $(\epsilon_q + \epsilon_{-q})$, remains finite, and thus the QFI $I_{\mathrm{NP}}(g_1)$ does not diverge. The scaling behaviors of $\hat{N}_1$ and the first excitation energy as functions of $\theta$ normalized by $\theta_{c}$ are presented in Fig. \ref{Fig:fig9}.
	
\end{widetext}

		\bibliography{triangle}

\begin{thebibliography}{47}%
\makeatletter
\providecommand \@ifxundefined [1]{%
 \@ifx{#1\undefined}
}%
\providecommand \@ifnum [1]{%
 \ifnum #1\expandafter \@firstoftwo
 \else \expandafter \@secondoftwo
 \fi
}%
\providecommand \@ifx [1]{%
 \ifx #1\expandafter \@firstoftwo
 \else \expandafter \@secondoftwo
 \fi
}%
\providecommand \natexlab [1]{#1}%
\providecommand \enquote  [1]{``#1''}%
\providecommand \bibnamefont  [1]{#1}%
\providecommand \bibfnamefont [1]{#1}%
\providecommand \citenamefont [1]{#1}%
\providecommand \href@noop [0]{\@secondoftwo}%
\providecommand \href [0]{\begingroup \@sanitize@url \@href}%
\providecommand \@href[1]{\@@startlink{#1}\@@href}%
\providecommand \@@href[1]{\endgroup#1\@@endlink}%
\providecommand \@sanitize@url [0]{\catcode `\\12\catcode `\$12\catcode
  `\&12\catcode `\#12\catcode `\^12\catcode `\_12\catcode `\%12\relax}%
\providecommand \@@startlink[1]{}%
\providecommand \@@endlink[0]{}%
\providecommand \url  [0]{\begingroup\@sanitize@url \@url }%
\providecommand \@url [1]{\endgroup\@href {#1}{\urlprefix }}%
\providecommand \urlprefix  [0]{URL }%
\providecommand \Eprint [0]{\href }%
\providecommand \doibase [0]{https://doi.org/}%
\providecommand \selectlanguage [0]{\@gobble}%
\providecommand \bibinfo  [0]{\@secondoftwo}%
\providecommand \bibfield  [0]{\@secondoftwo}%
\providecommand \translation [1]{[#1]}%
\providecommand \BibitemOpen [0]{}%
\providecommand \bibitemStop [0]{}%
\providecommand \bibitemNoStop [0]{.\EOS\space}%
\providecommand \EOS [0]{\spacefactor3000\relax}%
\providecommand \BibitemShut  [1]{\csname bibitem#1\endcsname}%
\let\auto@bib@innerbib\@empty
\bibitem [{\citenamefont {Degen}\ \emph {et~al.}(2017)\citenamefont {Degen},
  \citenamefont {Reinhard},\ and\ \citenamefont
  {Cappellaro}}]{RevModPhys.89.035002}%
  \BibitemOpen
  \bibfield  {author} {\bibinfo {author} {\bibfnamefont {C.~L.}\ \bibnamefont
  {Degen}}, \bibinfo {author} {\bibfnamefont {F.}~\bibnamefont {Reinhard}},\
  and\ \bibinfo {author} {\bibfnamefont {P.}~\bibnamefont {Cappellaro}},\
  }\bibfield  {title} {\bibinfo {title} {Quantum sensing},\ }\href
  {https://doi.org/10.1103/RevModPhys.89.035002} {\bibfield  {journal}
  {\bibinfo  {journal} {Rev. Mod. Phys.}\ }\textbf {\bibinfo {volume} {89}},\
  \bibinfo {pages} {035002} (\bibinfo {year} {2017})}\BibitemShut {NoStop}%
\bibitem [{\citenamefont {Demkowicz-Dobrza{\'n}ski}\ \emph
  {et~al.}(2015)\citenamefont {Demkowicz-Dobrza{\'n}ski}, \citenamefont
  {Jarzyna},\ and\ \citenamefont {Ko{\l}ody{\'n}ski}}]{demkowicz2015quantum}%
  \BibitemOpen
  \bibfield  {author} {\bibinfo {author} {\bibfnamefont {R.}~\bibnamefont
  {Demkowicz-Dobrza{\'n}ski}}, \bibinfo {author} {\bibfnamefont
  {M.}~\bibnamefont {Jarzyna}},\ and\ \bibinfo {author} {\bibfnamefont
  {J.}~\bibnamefont {Ko{\l}ody{\'n}ski}},\ }\bibfield  {title} {\bibinfo
  {title} {Quantum limits in optical interferometry},\ }\href@noop {}
  {\bibfield  {journal} {\bibinfo  {journal} {Progress in Optics}\ }\textbf
  {\bibinfo {volume} {60}},\ \bibinfo {pages} {345} (\bibinfo {year}
  {2015})}\BibitemShut {NoStop}%
\bibitem [{\citenamefont {Liu}\ \emph {et~al.}(2020)\citenamefont {Liu},
  \citenamefont {Yuan}, \citenamefont {Lu},\ and\ \citenamefont
  {Wang}}]{liu2020quantum}%
  \BibitemOpen
  \bibfield  {author} {\bibinfo {author} {\bibfnamefont {J.}~\bibnamefont
  {Liu}}, \bibinfo {author} {\bibfnamefont {H.}~\bibnamefont {Yuan}}, \bibinfo
  {author} {\bibfnamefont {X.-M.}\ \bibnamefont {Lu}},\ and\ \bibinfo {author}
  {\bibfnamefont {X.}~\bibnamefont {Wang}},\ }\bibfield  {title} {\bibinfo
  {title} {Quantum fisher information matrix and multiparameter estimation},\
  }\href@noop {} {\bibfield  {journal} {\bibinfo  {journal} {Journal of Physics
  A: Mathematical and Theoretical}\ }\textbf {\bibinfo {volume} {53}},\
  \bibinfo {pages} {023001} (\bibinfo {year} {2020})}\BibitemShut {NoStop}%
\bibitem [{\citenamefont {Giovannetti}\ \emph {et~al.}(2006)\citenamefont
  {Giovannetti}, \citenamefont {Lloyd},\ and\ \citenamefont
  {Maccone}}]{PhysRevLett.96.010401}%
  \BibitemOpen
  \bibfield  {author} {\bibinfo {author} {\bibfnamefont {V.}~\bibnamefont
  {Giovannetti}}, \bibinfo {author} {\bibfnamefont {S.}~\bibnamefont {Lloyd}},\
  and\ \bibinfo {author} {\bibfnamefont {L.}~\bibnamefont {Maccone}},\
  }\bibfield  {title} {\bibinfo {title} {Quantum metrology},\ }\href
  {https://doi.org/10.1103/PhysRevLett.96.010401} {\bibfield  {journal}
  {\bibinfo  {journal} {Phys. Rev. Lett.}\ }\textbf {\bibinfo {volume} {96}},\
  \bibinfo {pages} {010401} (\bibinfo {year} {2006})}\BibitemShut {NoStop}%
\bibitem [{\citenamefont {Giovannetti}\ \emph {et~al.}(2011)\citenamefont
  {Giovannetti}, \citenamefont {Lloyd},\ and\ \citenamefont
  {Maccone}}]{giovannetti2011advances}%
  \BibitemOpen
  \bibfield  {author} {\bibinfo {author} {\bibfnamefont {V.}~\bibnamefont
  {Giovannetti}}, \bibinfo {author} {\bibfnamefont {S.}~\bibnamefont {Lloyd}},\
  and\ \bibinfo {author} {\bibfnamefont {L.}~\bibnamefont {Maccone}},\
  }\bibfield  {title} {\bibinfo {title} {Advances in quantum metrology},\
  }\href@noop {} {\bibfield  {journal} {\bibinfo  {journal} {Nature photonics}\
  }\textbf {\bibinfo {volume} {5}},\ \bibinfo {pages} {222} (\bibinfo {year}
  {2011})}\BibitemShut {NoStop}%
\bibitem [{\citenamefont {Yuan}(2016)}]{PhysRevLett.117.160801}%
  \BibitemOpen
  \bibfield  {author} {\bibinfo {author} {\bibfnamefont {H.}~\bibnamefont
  {Yuan}},\ }\bibfield  {title} {\bibinfo {title} {Sequential feedback scheme
  outperforms the parallel scheme for hamiltonian parameter estimation},\
  }\href {https://doi.org/10.1103/PhysRevLett.117.160801} {\bibfield  {journal}
  {\bibinfo  {journal} {Phys. Rev. Lett.}\ }\textbf {\bibinfo {volume} {117}},\
  \bibinfo {pages} {160801} (\bibinfo {year} {2016})}\BibitemShut {NoStop}%
\bibitem [{\citenamefont {Pang}\ and\ \citenamefont
  {Jordan}(2017)}]{pang2017optimal}%
  \BibitemOpen
  \bibfield  {author} {\bibinfo {author} {\bibfnamefont {S.}~\bibnamefont
  {Pang}}\ and\ \bibinfo {author} {\bibfnamefont {A.~N.}\ \bibnamefont
  {Jordan}},\ }\bibfield  {title} {\bibinfo {title} {Optimal adaptive control
  for quantum metrology with time-dependent hamiltonians},\ }\href
  {https://doi.org/10.1038/ncomms14695} {\bibfield  {journal} {\bibinfo
  {journal} {Nature communications}\ }\textbf {\bibinfo {volume} {8}},\
  \bibinfo {pages} {14695} (\bibinfo {year} {2017})}\BibitemShut {NoStop}%
\bibitem [{\citenamefont {Hou}\ \emph {et~al.}(2021{\natexlab{a}})\citenamefont
  {Hou}, \citenamefont {Tang}, \citenamefont {Chen}, \citenamefont {Yuan},
  \citenamefont {Xiang}, \citenamefont {Li},\ and\ \citenamefont
  {Guo}}]{hou2021zero}%
  \BibitemOpen
  \bibfield  {author} {\bibinfo {author} {\bibfnamefont {Z.}~\bibnamefont
  {Hou}}, \bibinfo {author} {\bibfnamefont {J.-F.}\ \bibnamefont {Tang}},
  \bibinfo {author} {\bibfnamefont {H.}~\bibnamefont {Chen}}, \bibinfo {author}
  {\bibfnamefont {H.}~\bibnamefont {Yuan}}, \bibinfo {author} {\bibfnamefont
  {G.-Y.}\ \bibnamefont {Xiang}}, \bibinfo {author} {\bibfnamefont {C.-F.}\
  \bibnamefont {Li}},\ and\ \bibinfo {author} {\bibfnamefont {G.-C.}\
  \bibnamefont {Guo}},\ }\bibfield  {title} {\bibinfo {title} {Zero--trade-off
  multiparameter quantum estimation via simultaneously saturating multiple
  heisenberg uncertainty relations},\ }\href
  {https://doi.org/10.1126/sciadv.abd2986} {\bibfield  {journal} {\bibinfo
  {journal} {Science Advances}\ }\textbf {\bibinfo {volume} {7}},\ \bibinfo
  {pages} {eabd2986} (\bibinfo {year} {2021}{\natexlab{a}})}\BibitemShut
  {NoStop}%
\bibitem [{\citenamefont {Macieszczak}\ \emph {et~al.}(2016)\citenamefont
  {Macieszczak}, \citenamefont {Gu\ifmmode \mbox{\c{t}}\else
  \c{t}\fi{}\ifmmode~\u{a}\else \u{a}\fi{}}, \citenamefont {Lesanovsky},\ and\
  \citenamefont {Garrahan}}]{PhysRevA.93.022103}%
  \BibitemOpen
  \bibfield  {author} {\bibinfo {author} {\bibfnamefont {K.}~\bibnamefont
  {Macieszczak}}, \bibinfo {author} {\bibfnamefont {M.~u. u. u.~u.}\
  \bibnamefont {Gu\ifmmode \mbox{\c{t}}\else \c{t}\fi{}\ifmmode~\u{a}\else
  \u{a}\fi{}}}, \bibinfo {author} {\bibfnamefont {I.}~\bibnamefont
  {Lesanovsky}},\ and\ \bibinfo {author} {\bibfnamefont {J.~P.}\ \bibnamefont
  {Garrahan}},\ }\bibfield  {title} {\bibinfo {title} {Dynamical phase
  transitions as a resource for quantum enhanced metrology},\ }\href
  {https://doi.org/10.1103/PhysRevA.93.022103} {\bibfield  {journal} {\bibinfo
  {journal} {Phys. Rev. A}\ }\textbf {\bibinfo {volume} {93}},\ \bibinfo
  {pages} {022103} (\bibinfo {year} {2016})}\BibitemShut {NoStop}%
\bibitem [{\citenamefont {Braun}\ \emph {et~al.}(2018)\citenamefont {Braun},
  \citenamefont {Adesso}, \citenamefont {Benatti}, \citenamefont {Floreanini},
  \citenamefont {Marzolino}, \citenamefont {Mitchell},\ and\ \citenamefont
  {Pirandola}}]{RevModPhys.90.035006}%
  \BibitemOpen
  \bibfield  {author} {\bibinfo {author} {\bibfnamefont {D.}~\bibnamefont
  {Braun}}, \bibinfo {author} {\bibfnamefont {G.}~\bibnamefont {Adesso}},
  \bibinfo {author} {\bibfnamefont {F.}~\bibnamefont {Benatti}}, \bibinfo
  {author} {\bibfnamefont {R.}~\bibnamefont {Floreanini}}, \bibinfo {author}
  {\bibfnamefont {U.}~\bibnamefont {Marzolino}}, \bibinfo {author}
  {\bibfnamefont {M.~W.}\ \bibnamefont {Mitchell}},\ and\ \bibinfo {author}
  {\bibfnamefont {S.}~\bibnamefont {Pirandola}},\ }\bibfield  {title} {\bibinfo
  {title} {Quantum-enhanced measurements without entanglement},\ }\href
  {https://doi.org/10.1103/RevModPhys.90.035006} {\bibfield  {journal}
  {\bibinfo  {journal} {Rev. Mod. Phys.}\ }\textbf {\bibinfo {volume} {90}},\
  \bibinfo {pages} {035006} (\bibinfo {year} {2018})}\BibitemShut {NoStop}%
\bibitem [{\citenamefont {Ding}\ \emph {et~al.}(2022)\citenamefont {Ding},
  \citenamefont {Liu}, \citenamefont {Shi}, \citenamefont {Guo}, \citenamefont
  {M{\o}lmer},\ and\ \citenamefont {Adams}}]{ding2022enhanced}%
  \BibitemOpen
  \bibfield  {author} {\bibinfo {author} {\bibfnamefont {D.-S.}\ \bibnamefont
  {Ding}}, \bibinfo {author} {\bibfnamefont {Z.-K.}\ \bibnamefont {Liu}},
  \bibinfo {author} {\bibfnamefont {B.-S.}\ \bibnamefont {Shi}}, \bibinfo
  {author} {\bibfnamefont {G.-C.}\ \bibnamefont {Guo}}, \bibinfo {author}
  {\bibfnamefont {K.}~\bibnamefont {M{\o}lmer}},\ and\ \bibinfo {author}
  {\bibfnamefont {C.~S.}\ \bibnamefont {Adams}},\ }\bibfield  {title} {\bibinfo
  {title} {Enhanced metrology at the critical point of a many-body rydberg
  atomic system},\ }\href {https://doi.org/10.1038/s41567-022-01777-8}
  {\bibfield  {journal} {\bibinfo  {journal} {Nature Physics}\ }\textbf
  {\bibinfo {volume} {18}},\ \bibinfo {pages} {1447} (\bibinfo {year}
  {2022})}\BibitemShut {NoStop}%
\bibitem [{\citenamefont {Hotter}\ \emph {et~al.}(2024)\citenamefont {Hotter},
  \citenamefont {Ritsch},\ and\ \citenamefont
  {Gietka}}]{PhysRevLett.132.060801}%
  \BibitemOpen
  \bibfield  {author} {\bibinfo {author} {\bibfnamefont {C.}~\bibnamefont
  {Hotter}}, \bibinfo {author} {\bibfnamefont {H.}~\bibnamefont {Ritsch}},\
  and\ \bibinfo {author} {\bibfnamefont {K.}~\bibnamefont {Gietka}},\
  }\bibfield  {title} {\bibinfo {title} {Combining critical and quantum
  metrology},\ }\href {https://doi.org/10.1103/PhysRevLett.132.060801}
  {\bibfield  {journal} {\bibinfo  {journal} {Phys. Rev. Lett.}\ }\textbf
  {\bibinfo {volume} {132}},\ \bibinfo {pages} {060801} (\bibinfo {year}
  {2024})}\BibitemShut {NoStop}%
\bibitem [{\citenamefont {Yang}\ \emph {et~al.}(2024)\citenamefont {Yang},
  \citenamefont {Yuan},\ and\ \citenamefont {Li}}]{PhysRevA.109.022604}%
  \BibitemOpen
  \bibfield  {author} {\bibinfo {author} {\bibfnamefont {Y.}~\bibnamefont
  {Yang}}, \bibinfo {author} {\bibfnamefont {H.}~\bibnamefont {Yuan}},\ and\
  \bibinfo {author} {\bibfnamefont {F.}~\bibnamefont {Li}},\ }\bibfield
  {title} {\bibinfo {title} {Quantum multiparameter estimation enhanced by a
  topological phase transition},\ }\href
  {https://doi.org/10.1103/PhysRevA.109.022604} {\bibfield  {journal} {\bibinfo
   {journal} {Phys. Rev. A}\ }\textbf {\bibinfo {volume} {109}},\ \bibinfo
  {pages} {022604} (\bibinfo {year} {2024})}\BibitemShut {NoStop}%
\bibitem [{\citenamefont {Yang}(2007)}]{PhysRevB.76.180403}%
  \BibitemOpen
  \bibfield  {author} {\bibinfo {author} {\bibfnamefont {M.-F.}\ \bibnamefont
  {Yang}},\ }\bibfield  {title} {\bibinfo {title} {Ground-state fidelity in
  one-dimensional gapless models},\ }\href
  {https://doi.org/10.1103/PhysRevB.76.180403} {\bibfield  {journal} {\bibinfo
  {journal} {Phys. Rev. B}\ }\textbf {\bibinfo {volume} {76}},\ \bibinfo
  {pages} {180403} (\bibinfo {year} {2007})}\BibitemShut {NoStop}%
\bibitem [{\citenamefont {Ilias}\ \emph {et~al.}(2022)\citenamefont {Ilias},
  \citenamefont {Yang}, \citenamefont {Huelga},\ and\ \citenamefont
  {Plenio}}]{PRXQuantum.3.010354}%
  \BibitemOpen
  \bibfield  {author} {\bibinfo {author} {\bibfnamefont {T.}~\bibnamefont
  {Ilias}}, \bibinfo {author} {\bibfnamefont {D.}~\bibnamefont {Yang}},
  \bibinfo {author} {\bibfnamefont {S.~F.}\ \bibnamefont {Huelga}},\ and\
  \bibinfo {author} {\bibfnamefont {M.~B.}\ \bibnamefont {Plenio}},\ }\bibfield
   {title} {\bibinfo {title} {Criticality-enhanced quantum sensing via
  continuous measurement},\ }\href
  {https://doi.org/10.1103/PRXQuantum.3.010354} {\bibfield  {journal} {\bibinfo
   {journal} {PRX Quantum}\ }\textbf {\bibinfo {volume} {3}},\ \bibinfo {pages}
  {010354} (\bibinfo {year} {2022})}\BibitemShut {NoStop}%
\bibitem [{\citenamefont {Chu}\ \emph {et~al.}(2021)\citenamefont {Chu},
  \citenamefont {Zhang}, \citenamefont {Yu},\ and\ \citenamefont
  {Cai}}]{PhysRevLett.126.010502}%
  \BibitemOpen
  \bibfield  {author} {\bibinfo {author} {\bibfnamefont {Y.}~\bibnamefont
  {Chu}}, \bibinfo {author} {\bibfnamefont {S.}~\bibnamefont {Zhang}}, \bibinfo
  {author} {\bibfnamefont {B.}~\bibnamefont {Yu}},\ and\ \bibinfo {author}
  {\bibfnamefont {J.}~\bibnamefont {Cai}},\ }\bibfield  {title} {\bibinfo
  {title} {Dynamic framework for criticality-enhanced quantum sensing},\ }\href
  {https://doi.org/10.1103/PhysRevLett.126.010502} {\bibfield  {journal}
  {\bibinfo  {journal} {Phys. Rev. Lett.}\ }\textbf {\bibinfo {volume} {126}},\
  \bibinfo {pages} {010502} (\bibinfo {year} {2021})}\BibitemShut {NoStop}%
\bibitem [{\citenamefont {Zhang}\ \emph {et~al.}(2024)\citenamefont {Zhang},
  \citenamefont {Ding}, \citenamefont {Zhang}, \citenamefont {Shao},
  \citenamefont {Zhang},\ and\ \citenamefont {Wang}}]{PhysRevA.110.012413}%
  \BibitemOpen
  \bibfield  {author} {\bibinfo {author} {\bibfnamefont {R.}~\bibnamefont
  {Zhang}}, \bibinfo {author} {\bibfnamefont {W.}~\bibnamefont {Ding}},
  \bibinfo {author} {\bibfnamefont {Z.}~\bibnamefont {Zhang}}, \bibinfo
  {author} {\bibfnamefont {L.}~\bibnamefont {Shao}}, \bibinfo {author}
  {\bibfnamefont {Y.}~\bibnamefont {Zhang}},\ and\ \bibinfo {author}
  {\bibfnamefont {X.}~\bibnamefont {Wang}},\ }\bibfield  {title} {\bibinfo
  {title} {Relations between quantum metrology and criticality in general
  su(1,1) systems},\ }\href {https://doi.org/10.1103/PhysRevA.110.012413}
  {\bibfield  {journal} {\bibinfo  {journal} {Phys. Rev. A}\ }\textbf {\bibinfo
  {volume} {110}},\ \bibinfo {pages} {012413} (\bibinfo {year}
  {2024})}\BibitemShut {NoStop}%
\bibitem [{\citenamefont {Tang}\ \emph {et~al.}(2024)\citenamefont {Tang},
  \citenamefont {Yang}, \citenamefont {An}, \citenamefont {Xie}, \citenamefont
  {Wang}, \citenamefont {Wang},\ and\ \citenamefont
  {Li}}]{PhysRevA.110.022611}%
  \BibitemOpen
  \bibfield  {author} {\bibinfo {author} {\bibfnamefont {Y.}~\bibnamefont
  {Tang}}, \bibinfo {author} {\bibfnamefont {Y.}~\bibnamefont {Yang}}, \bibinfo
  {author} {\bibfnamefont {M.}~\bibnamefont {An}}, \bibinfo {author}
  {\bibfnamefont {J.}~\bibnamefont {Xie}}, \bibinfo {author} {\bibfnamefont
  {Y.}~\bibnamefont {Wang}}, \bibinfo {author} {\bibfnamefont {F.}~\bibnamefont
  {Wang}},\ and\ \bibinfo {author} {\bibfnamefont {F.}~\bibnamefont {Li}},\
  }\bibfield  {title} {\bibinfo {title} {Critical properties of quantum fisher
  information of su(1,1)-dynamic systems},\ }\href
  {https://doi.org/10.1103/PhysRevA.110.022611} {\bibfield  {journal} {\bibinfo
   {journal} {Phys. Rev. A}\ }\textbf {\bibinfo {volume} {110}},\ \bibinfo
  {pages} {022611} (\bibinfo {year} {2024})}\BibitemShut {NoStop}%
\bibitem [{\citenamefont {Ma}\ and\ \citenamefont
  {Wang}(2009)}]{PhysRevA.80.012318}%
  \BibitemOpen
  \bibfield  {author} {\bibinfo {author} {\bibfnamefont {J.}~\bibnamefont
  {Ma}}\ and\ \bibinfo {author} {\bibfnamefont {X.}~\bibnamefont {Wang}},\
  }\bibfield  {title} {\bibinfo {title} {Fisher information and spin squeezing
  in the lipkin-meshkov-glick model},\ }\href
  {https://doi.org/10.1103/PhysRevA.80.012318} {\bibfield  {journal} {\bibinfo
  {journal} {Phys. Rev. A}\ }\textbf {\bibinfo {volume} {80}},\ \bibinfo
  {pages} {012318} (\bibinfo {year} {2009})}\BibitemShut {NoStop}%
\bibitem [{\citenamefont {Salvatori}\ \emph {et~al.}(2014)\citenamefont
  {Salvatori}, \citenamefont {Mandarino},\ and\ \citenamefont
  {Paris}}]{PhysRevA.90.022111}%
  \BibitemOpen
  \bibfield  {author} {\bibinfo {author} {\bibfnamefont {G.}~\bibnamefont
  {Salvatori}}, \bibinfo {author} {\bibfnamefont {A.}~\bibnamefont
  {Mandarino}},\ and\ \bibinfo {author} {\bibfnamefont {M.~G.~A.}\ \bibnamefont
  {Paris}},\ }\bibfield  {title} {\bibinfo {title} {Quantum metrology in
  lipkin-meshkov-glick critical systems},\ }\href
  {https://doi.org/10.1103/PhysRevA.90.022111} {\bibfield  {journal} {\bibinfo
  {journal} {Phys. Rev. A}\ }\textbf {\bibinfo {volume} {90}},\ \bibinfo
  {pages} {022111} (\bibinfo {year} {2014})}\BibitemShut {NoStop}%
\bibitem [{\citenamefont {Carollo}\ \emph {et~al.}(2020)\citenamefont
  {Carollo}, \citenamefont {Valenti},\ and\ \citenamefont
  {Spagnolo}}]{CAROLLO20201}%
  \BibitemOpen
  \bibfield  {author} {\bibinfo {author} {\bibfnamefont {A.}~\bibnamefont
  {Carollo}}, \bibinfo {author} {\bibfnamefont {D.}~\bibnamefont {Valenti}},\
  and\ \bibinfo {author} {\bibfnamefont {B.}~\bibnamefont {Spagnolo}},\
  }\bibfield  {title} {\bibinfo {title} {Geometry of quantum phase
  transitions},\ }\href
  {https://doi.org/https://doi.org/10.1016/j.physrep.2019.11.002} {\bibfield
  {journal} {\bibinfo  {journal} {Physics Reports}\ }\textbf {\bibinfo {volume}
  {838}},\ \bibinfo {pages} {1} (\bibinfo {year} {2020})},\ \bibinfo {note}
  {geometry of quantum phase transitions}\BibitemShut {NoStop}%
\bibitem [{\citenamefont {Montenegro}\ \emph {et~al.}(2021)\citenamefont
  {Montenegro}, \citenamefont {Mishra},\ and\ \citenamefont
  {Bayat}}]{PhysRevLett.126.200501}%
  \BibitemOpen
  \bibfield  {author} {\bibinfo {author} {\bibfnamefont {V.}~\bibnamefont
  {Montenegro}}, \bibinfo {author} {\bibfnamefont {U.}~\bibnamefont {Mishra}},\
  and\ \bibinfo {author} {\bibfnamefont {A.}~\bibnamefont {Bayat}},\ }\bibfield
   {title} {\bibinfo {title} {Global sensing and its impact for quantum
  many-body probes with criticality},\ }\href
  {https://doi.org/10.1103/PhysRevLett.126.200501} {\bibfield  {journal}
  {\bibinfo  {journal} {Phys. Rev. Lett.}\ }\textbf {\bibinfo {volume} {126}},\
  \bibinfo {pages} {200501} (\bibinfo {year} {2021})}\BibitemShut {NoStop}%
\bibitem [{\citenamefont {Mukhopadhyay}\ and\ \citenamefont
  {Bayat}(2024)}]{PhysRevLett.133.120601}%
  \BibitemOpen
  \bibfield  {author} {\bibinfo {author} {\bibfnamefont {C.}~\bibnamefont
  {Mukhopadhyay}}\ and\ \bibinfo {author} {\bibfnamefont {A.}~\bibnamefont
  {Bayat}},\ }\bibfield  {title} {\bibinfo {title} {Modular many-body quantum
  sensors},\ }\href {https://doi.org/10.1103/PhysRevLett.133.120601} {\bibfield
   {journal} {\bibinfo  {journal} {Phys. Rev. Lett.}\ }\textbf {\bibinfo
  {volume} {133}},\ \bibinfo {pages} {120601} (\bibinfo {year}
  {2024})}\BibitemShut {NoStop}%
\bibitem [{\citenamefont {Cheng}\ \emph {et~al.}(2025)\citenamefont {Cheng},
  \citenamefont {Zhang}, \citenamefont {Zhou},\ and\ \citenamefont
  {Zhou}}]{PhysRevLett.134.190802}%
  \BibitemOpen
  \bibfield  {author} {\bibinfo {author} {\bibfnamefont {J.-M.}\ \bibnamefont
  {Cheng}}, \bibinfo {author} {\bibfnamefont {Y.-C.}\ \bibnamefont {Zhang}},
  \bibinfo {author} {\bibfnamefont {X.-F.}\ \bibnamefont {Zhou}},\ and\
  \bibinfo {author} {\bibfnamefont {Z.-W.}\ \bibnamefont {Zhou}},\ }\bibfield
  {title} {\bibinfo {title} {Super-heisenberg scaling in a triple-point
  criticality},\ }\href {https://doi.org/10.1103/PhysRevLett.134.190802}
  {\bibfield  {journal} {\bibinfo  {journal} {Phys. Rev. Lett.}\ }\textbf
  {\bibinfo {volume} {134}},\ \bibinfo {pages} {190802} (\bibinfo {year}
  {2025})}\BibitemShut {NoStop}%
\bibitem [{\citenamefont {Zhang}\ \emph {et~al.}(2021)\citenamefont {Zhang},
  \citenamefont {Hu}, \citenamefont {Fu}, \citenamefont {Luo}, \citenamefont
  {Pu},\ and\ \citenamefont {Zhang}}]{PhysRevLett.127.063602}%
  \BibitemOpen
  \bibfield  {author} {\bibinfo {author} {\bibfnamefont {Y.-Y.}\ \bibnamefont
  {Zhang}}, \bibinfo {author} {\bibfnamefont {Z.-X.}\ \bibnamefont {Hu}},
  \bibinfo {author} {\bibfnamefont {L.}~\bibnamefont {Fu}}, \bibinfo {author}
  {\bibfnamefont {H.-G.}\ \bibnamefont {Luo}}, \bibinfo {author} {\bibfnamefont
  {H.}~\bibnamefont {Pu}},\ and\ \bibinfo {author} {\bibfnamefont {X.-F.}\
  \bibnamefont {Zhang}},\ }\bibfield  {title} {\bibinfo {title} {Quantum phases
  in a quantum rabi triangle},\ }\href
  {https://doi.org/10.1103/PhysRevLett.127.063602} {\bibfield  {journal}
  {\bibinfo  {journal} {Phys. Rev. Lett.}\ }\textbf {\bibinfo {volume} {127}},\
  \bibinfo {pages} {063602} (\bibinfo {year} {2021})}\BibitemShut {NoStop}%
\bibitem [{\citenamefont {Fallas~Padilla}\ \emph {et~al.}(2022)\citenamefont
  {Fallas~Padilla}, \citenamefont {Pu}, \citenamefont {Cheng},\ and\
  \citenamefont {Zhang}}]{PhysRevLett.129.183602}%
  \BibitemOpen
  \bibfield  {author} {\bibinfo {author} {\bibfnamefont {D.}~\bibnamefont
  {Fallas~Padilla}}, \bibinfo {author} {\bibfnamefont {H.}~\bibnamefont {Pu}},
  \bibinfo {author} {\bibfnamefont {G.-J.}\ \bibnamefont {Cheng}},\ and\
  \bibinfo {author} {\bibfnamefont {Y.-Y.}\ \bibnamefont {Zhang}},\ }\bibfield
  {title} {\bibinfo {title} {Understanding the quantum rabi ring using
  analogies to quantum magnetism},\ }\href
  {https://doi.org/10.1103/PhysRevLett.129.183602} {\bibfield  {journal}
  {\bibinfo  {journal} {Phys. Rev. Lett.}\ }\textbf {\bibinfo {volume} {129}},\
  \bibinfo {pages} {183602} (\bibinfo {year} {2022})}\BibitemShut {NoStop}%
\bibitem [{\citenamefont {Bakemeier}\ \emph {et~al.}(2012)\citenamefont
  {Bakemeier}, \citenamefont {Alvermann},\ and\ \citenamefont
  {Fehske}}]{PhysRevA.85.043821}%
  \BibitemOpen
  \bibfield  {author} {\bibinfo {author} {\bibfnamefont {L.}~\bibnamefont
  {Bakemeier}}, \bibinfo {author} {\bibfnamefont {A.}~\bibnamefont
  {Alvermann}},\ and\ \bibinfo {author} {\bibfnamefont {H.}~\bibnamefont
  {Fehske}},\ }\bibfield  {title} {\bibinfo {title} {Quantum phase transition
  in the dicke model with critical and noncritical entanglement},\ }\href
  {https://doi.org/10.1103/PhysRevA.85.043821} {\bibfield  {journal} {\bibinfo
  {journal} {Phys. Rev. A}\ }\textbf {\bibinfo {volume} {85}},\ \bibinfo
  {pages} {043821} (\bibinfo {year} {2012})}\BibitemShut {NoStop}%
\bibitem [{\citenamefont {Hwang}\ \emph {et~al.}(2015)\citenamefont {Hwang},
  \citenamefont {Puebla},\ and\ \citenamefont {Plenio}}]{hwang2015quantum}%
  \BibitemOpen
  \bibfield  {author} {\bibinfo {author} {\bibfnamefont {M.-J.}\ \bibnamefont
  {Hwang}}, \bibinfo {author} {\bibfnamefont {R.}~\bibnamefont {Puebla}},\ and\
  \bibinfo {author} {\bibfnamefont {M.~B.}\ \bibnamefont {Plenio}},\ }\bibfield
   {title} {\bibinfo {title} {Quantum phase transition and universal dynamics
  in the rabi model},\ }\href {https://doi.org/10.1103/PhysRevLett.115.180404}
  {\bibfield  {journal} {\bibinfo  {journal} {Phys. Rev. Lett.}\ }\textbf
  {\bibinfo {volume} {115}},\ \bibinfo {pages} {180404} (\bibinfo {year}
  {2015})}\BibitemShut {NoStop}%
\bibitem [{\citenamefont {Qin}\ and\ \citenamefont
  {Zhang}(2024)}]{PhysRevA.110.013713}%
  \BibitemOpen
  \bibfield  {author} {\bibinfo {author} {\bibfnamefont {X.}~\bibnamefont
  {Qin}}\ and\ \bibinfo {author} {\bibfnamefont {Y.-Y.}\ \bibnamefont
  {Zhang}},\ }\bibfield  {title} {\bibinfo {title} {Quantum fluctuations and
  unusual critical exponents in a quantum rabi triangle},\ }\href
  {https://doi.org/10.1103/PhysRevA.110.013713} {\bibfield  {journal} {\bibinfo
   {journal} {Phys. Rev. A}\ }\textbf {\bibinfo {volume} {110}},\ \bibinfo
  {pages} {013713} (\bibinfo {year} {2024})}\BibitemShut {NoStop}%
\bibitem [{\citenamefont {Manucharyan}\ \emph {et~al.}(2017)\citenamefont
  {Manucharyan}, \citenamefont {Baksic},\ and\ \citenamefont
  {Ciuti}}]{manucharyan2017resilience}%
  \BibitemOpen
  \bibfield  {author} {\bibinfo {author} {\bibfnamefont {V.~E.}\ \bibnamefont
  {Manucharyan}}, \bibinfo {author} {\bibfnamefont {A.}~\bibnamefont
  {Baksic}},\ and\ \bibinfo {author} {\bibfnamefont {C.}~\bibnamefont
  {Ciuti}},\ }\bibfield  {title} {\bibinfo {title} {Resilience of the quantum
  rabi model in circuit qed},\ }\href@noop {} {\bibfield  {journal} {\bibinfo
  {journal} {Journal of Physics A: Mathematical and Theoretical}\ }\textbf
  {\bibinfo {volume} {50}},\ \bibinfo {pages} {294001} (\bibinfo {year}
  {2017})}\BibitemShut {NoStop}%
\bibitem [{\citenamefont {Gely}\ \emph {et~al.}(2017)\citenamefont {Gely},
  \citenamefont {Parra-Rodriguez}, \citenamefont {Bothner}, \citenamefont
  {Blanter}, \citenamefont {Bosman}, \citenamefont {Solano},\ and\
  \citenamefont {Steele}}]{PhysRevB.95.245115}%
  \BibitemOpen
  \bibfield  {author} {\bibinfo {author} {\bibfnamefont {M.~F.}\ \bibnamefont
  {Gely}}, \bibinfo {author} {\bibfnamefont {A.}~\bibnamefont
  {Parra-Rodriguez}}, \bibinfo {author} {\bibfnamefont {D.}~\bibnamefont
  {Bothner}}, \bibinfo {author} {\bibfnamefont {Y.~M.}\ \bibnamefont
  {Blanter}}, \bibinfo {author} {\bibfnamefont {S.~J.}\ \bibnamefont {Bosman}},
  \bibinfo {author} {\bibfnamefont {E.}~\bibnamefont {Solano}},\ and\ \bibinfo
  {author} {\bibfnamefont {G.~A.}\ \bibnamefont {Steele}},\ }\bibfield  {title}
  {\bibinfo {title} {Convergence of the multimode quantum rabi model of circuit
  quantum electrodynamics},\ }\href
  {https://doi.org/10.1103/PhysRevB.95.245115} {\bibfield  {journal} {\bibinfo
  {journal} {Phys. Rev. B}\ }\textbf {\bibinfo {volume} {95}},\ \bibinfo
  {pages} {245115} (\bibinfo {year} {2017})}\BibitemShut {NoStop}%
\bibitem [{\citenamefont {Felicetti}\ \emph {et~al.}(2018)\citenamefont
  {Felicetti}, \citenamefont {Rossatto}, \citenamefont {Rico}, \citenamefont
  {Solano},\ and\ \citenamefont {Forn-D\'{\i}az}}]{PhysRevA.97.013851}%
  \BibitemOpen
  \bibfield  {author} {\bibinfo {author} {\bibfnamefont {S.}~\bibnamefont
  {Felicetti}}, \bibinfo {author} {\bibfnamefont {D.~Z.}\ \bibnamefont
  {Rossatto}}, \bibinfo {author} {\bibfnamefont {E.}~\bibnamefont {Rico}},
  \bibinfo {author} {\bibfnamefont {E.}~\bibnamefont {Solano}},\ and\ \bibinfo
  {author} {\bibfnamefont {P.}~\bibnamefont {Forn-D\'{\i}az}},\ }\bibfield
  {title} {\bibinfo {title} {Two-photon quantum rabi model with superconducting
  circuits},\ }\href {https://doi.org/10.1103/PhysRevA.97.013851} {\bibfield
  {journal} {\bibinfo  {journal} {Phys. Rev. A}\ }\textbf {\bibinfo {volume}
  {97}},\ \bibinfo {pages} {013851} (\bibinfo {year} {2018})}\BibitemShut
  {NoStop}%
\bibitem [{\citenamefont {Roth}\ \emph {et~al.}(2019)\citenamefont {Roth},
  \citenamefont {Hassler},\ and\ \citenamefont
  {DiVincenzo}}]{PhysRevResearch.1.033128}%
  \BibitemOpen
  \bibfield  {author} {\bibinfo {author} {\bibfnamefont {M.}~\bibnamefont
  {Roth}}, \bibinfo {author} {\bibfnamefont {F.}~\bibnamefont {Hassler}},\ and\
  \bibinfo {author} {\bibfnamefont {D.~P.}\ \bibnamefont {DiVincenzo}},\
  }\bibfield  {title} {\bibinfo {title} {Optimal gauge for the multimode rabi
  model in circuit qed},\ }\href
  {https://doi.org/10.1103/PhysRevResearch.1.033128} {\bibfield  {journal}
  {\bibinfo  {journal} {Phys. Rev. Res.}\ }\textbf {\bibinfo {volume} {1}},\
  \bibinfo {pages} {033128} (\bibinfo {year} {2019})}\BibitemShut {NoStop}%
\bibitem [{\citenamefont {Polino}\ \emph {et~al.}(2020)\citenamefont {Polino},
  \citenamefont {Valeri}, \citenamefont {Spagnolo},\ and\ \citenamefont
  {Sciarrino}}]{polino2020photonic}%
  \BibitemOpen
  \bibfield  {author} {\bibinfo {author} {\bibfnamefont {E.}~\bibnamefont
  {Polino}}, \bibinfo {author} {\bibfnamefont {M.}~\bibnamefont {Valeri}},
  \bibinfo {author} {\bibfnamefont {N.}~\bibnamefont {Spagnolo}},\ and\
  \bibinfo {author} {\bibfnamefont {F.}~\bibnamefont {Sciarrino}},\ }\bibfield
  {title} {\bibinfo {title} {Photonic quantum metrology},\ }\href@noop {}
  {\bibfield  {journal} {\bibinfo  {journal} {AVS Quantum Science}\ }\textbf
  {\bibinfo {volume} {2}} (\bibinfo {year} {2020})}\BibitemShut {NoStop}%
\bibitem [{\citenamefont {Gibbons}(1992)}]{GIBBONS1992147}%
  \BibitemOpen
  \bibfield  {author} {\bibinfo {author} {\bibfnamefont {G.}~\bibnamefont
  {Gibbons}},\ }\bibfield  {title} {\bibinfo {title} {Typical states and
  density matrices},\ }\href
  {https://doi.org/https://doi.org/10.1016/0393-0440(92)90046-4} {\bibfield
  {journal} {\bibinfo  {journal} {Journal of Geometry and Physics}\ }\textbf
  {\bibinfo {volume} {8}},\ \bibinfo {pages} {147} (\bibinfo {year}
  {1992})}\BibitemShut {NoStop}%
\bibitem [{\citenamefont {Braunstein}\ and\ \citenamefont
  {Caves}(1994)}]{PhysRevLett.72.3439}%
  \BibitemOpen
  \bibfield  {author} {\bibinfo {author} {\bibfnamefont {S.~L.}\ \bibnamefont
  {Braunstein}}\ and\ \bibinfo {author} {\bibfnamefont {C.~M.}\ \bibnamefont
  {Caves}},\ }\bibfield  {title} {\bibinfo {title} {Statistical distance and
  the geometry of quantum states},\ }\href
  {https://doi.org/10.1103/PhysRevLett.72.3439} {\bibfield  {journal} {\bibinfo
   {journal} {Phys. Rev. Lett.}\ }\textbf {\bibinfo {volume} {72}},\ \bibinfo
  {pages} {3439} (\bibinfo {year} {1994})}\BibitemShut {NoStop}%
\bibitem [{\citenamefont {Zanardi}\ \emph {et~al.}(2007)\citenamefont
  {Zanardi}, \citenamefont {Giorda},\ and\ \citenamefont
  {Cozzini}}]{PhysRevLett.99.100603}%
  \BibitemOpen
  \bibfield  {author} {\bibinfo {author} {\bibfnamefont {P.}~\bibnamefont
  {Zanardi}}, \bibinfo {author} {\bibfnamefont {P.}~\bibnamefont {Giorda}},\
  and\ \bibinfo {author} {\bibfnamefont {M.}~\bibnamefont {Cozzini}},\
  }\bibfield  {title} {\bibinfo {title} {Information-theoretic differential
  geometry of quantum phase transitions},\ }\href
  {https://doi.org/10.1103/PhysRevLett.99.100603} {\bibfield  {journal}
  {\bibinfo  {journal} {Phys. Rev. Lett.}\ }\textbf {\bibinfo {volume} {99}},\
  \bibinfo {pages} {100603} (\bibinfo {year} {2007})}\BibitemShut {NoStop}%
\bibitem [{\citenamefont {Sakurai}\ and\ \citenamefont
  {Napolitano}(2020)}]{sakurai2020modern}%
  \BibitemOpen
  \bibfield  {author} {\bibinfo {author} {\bibfnamefont {J.~J.}\ \bibnamefont
  {Sakurai}}\ and\ \bibinfo {author} {\bibfnamefont {J.}~\bibnamefont
  {Napolitano}},\ }\href@noop {} {\emph {\bibinfo {title} {Modern quantum
  mechanics}}}\ (\bibinfo  {publisher} {Cambridge University Press},\ \bibinfo
  {year} {2020})\BibitemShut {NoStop}%
\bibitem [{\citenamefont {Fisher}(1967)}]{fisher1967theory}%
  \BibitemOpen
  \bibfield  {author} {\bibinfo {author} {\bibfnamefont {M.~E.}\ \bibnamefont
  {Fisher}},\ }\bibfield  {title} {\bibinfo {title} {The theory of equilibrium
  critical phenomena},\ }\href@noop {} {\bibfield  {journal} {\bibinfo
  {journal} {Reports on progress in physics}\ }\textbf {\bibinfo {volume}
  {30}},\ \bibinfo {pages} {615} (\bibinfo {year} {1967})}\BibitemShut
  {NoStop}%
\bibitem [{\citenamefont {Hohenberg}\ and\ \citenamefont
  {Halperin}(1977)}]{RevModPhys.49.435}%
  \BibitemOpen
  \bibfield  {author} {\bibinfo {author} {\bibfnamefont {P.~C.}\ \bibnamefont
  {Hohenberg}}\ and\ \bibinfo {author} {\bibfnamefont {B.~I.}\ \bibnamefont
  {Halperin}},\ }\bibfield  {title} {\bibinfo {title} {Theory of dynamic
  critical phenomena},\ }\href {https://doi.org/10.1103/RevModPhys.49.435}
  {\bibfield  {journal} {\bibinfo  {journal} {Rev. Mod. Phys.}\ }\textbf
  {\bibinfo {volume} {49}},\ \bibinfo {pages} {435} (\bibinfo {year}
  {1977})}\BibitemShut {NoStop}%
\bibitem [{\citenamefont {Giovannetti}\ \emph {et~al.}(2004)\citenamefont
  {Giovannetti}, \citenamefont {Lloyd},\ and\ \citenamefont
  {Maccone}}]{giovannetti2004quantum}%
  \BibitemOpen
  \bibfield  {author} {\bibinfo {author} {\bibfnamefont {V.}~\bibnamefont
  {Giovannetti}}, \bibinfo {author} {\bibfnamefont {S.}~\bibnamefont {Lloyd}},\
  and\ \bibinfo {author} {\bibfnamefont {L.}~\bibnamefont {Maccone}},\
  }\bibfield  {title} {\bibinfo {title} {Quantum-enhanced measurements: beating
  the standard quantum limit},\ }\href@noop {} {\bibfield  {journal} {\bibinfo
  {journal} {Science}\ }\textbf {\bibinfo {volume} {306}},\ \bibinfo {pages}
  {1330} (\bibinfo {year} {2004})}\BibitemShut {NoStop}%
\bibitem [{\citenamefont {Boixo}\ \emph {et~al.}(2007)\citenamefont {Boixo},
  \citenamefont {Flammia}, \citenamefont {Caves},\ and\ \citenamefont
  {Geremia}}]{PhysRevLett.98.090401}%
  \BibitemOpen
  \bibfield  {author} {\bibinfo {author} {\bibfnamefont {S.}~\bibnamefont
  {Boixo}}, \bibinfo {author} {\bibfnamefont {S.~T.}\ \bibnamefont {Flammia}},
  \bibinfo {author} {\bibfnamefont {C.~M.}\ \bibnamefont {Caves}},\ and\
  \bibinfo {author} {\bibfnamefont {J.}~\bibnamefont {Geremia}},\ }\bibfield
  {title} {\bibinfo {title} {Generalized limits for single-parameter quantum
  estimation},\ }\href {https://doi.org/10.1103/PhysRevLett.98.090401}
  {\bibfield  {journal} {\bibinfo  {journal} {Phys. Rev. Lett.}\ }\textbf
  {\bibinfo {volume} {98}},\ \bibinfo {pages} {090401} (\bibinfo {year}
  {2007})}\BibitemShut {NoStop}%
\bibitem [{\citenamefont {Hou}\ \emph {et~al.}(2021{\natexlab{b}})\citenamefont
  {Hou}, \citenamefont {Jin}, \citenamefont {Chen}, \citenamefont {Tang},
  \citenamefont {Huang}, \citenamefont {Yuan}, \citenamefont {Xiang},
  \citenamefont {Li},\ and\ \citenamefont {Guo}}]{PhysRevLett.126.070503}%
  \BibitemOpen
  \bibfield  {author} {\bibinfo {author} {\bibfnamefont {Z.}~\bibnamefont
  {Hou}}, \bibinfo {author} {\bibfnamefont {Y.}~\bibnamefont {Jin}}, \bibinfo
  {author} {\bibfnamefont {H.}~\bibnamefont {Chen}}, \bibinfo {author}
  {\bibfnamefont {J.-F.}\ \bibnamefont {Tang}}, \bibinfo {author}
  {\bibfnamefont {C.-J.}\ \bibnamefont {Huang}}, \bibinfo {author}
  {\bibfnamefont {H.}~\bibnamefont {Yuan}}, \bibinfo {author} {\bibfnamefont
  {G.-Y.}\ \bibnamefont {Xiang}}, \bibinfo {author} {\bibfnamefont {C.-F.}\
  \bibnamefont {Li}},\ and\ \bibinfo {author} {\bibfnamefont {G.-C.}\
  \bibnamefont {Guo}},\ }\bibfield  {title} {\bibinfo {title}
  {``super-heisenberg'' and heisenberg scalings achieved simultaneously in the
  estimation of a rotating field},\ }\href
  {https://doi.org/10.1103/PhysRevLett.126.070503} {\bibfield  {journal}
  {\bibinfo  {journal} {Phys. Rev. Lett.}\ }\textbf {\bibinfo {volume} {126}},\
  \bibinfo {pages} {070503} (\bibinfo {year} {2021}{\natexlab{b}})}\BibitemShut
  {NoStop}%
\bibitem [{\citenamefont {Garbe}\ \emph {et~al.}(2020)\citenamefont {Garbe},
  \citenamefont {Bina}, \citenamefont {Keller}, \citenamefont {Paris},\ and\
  \citenamefont {Felicetti}}]{PhysRevLett.124.120504}%
  \BibitemOpen
  \bibfield  {author} {\bibinfo {author} {\bibfnamefont {L.}~\bibnamefont
  {Garbe}}, \bibinfo {author} {\bibfnamefont {M.}~\bibnamefont {Bina}},
  \bibinfo {author} {\bibfnamefont {A.}~\bibnamefont {Keller}}, \bibinfo
  {author} {\bibfnamefont {M.~G.~A.}\ \bibnamefont {Paris}},\ and\ \bibinfo
  {author} {\bibfnamefont {S.}~\bibnamefont {Felicetti}},\ }\bibfield  {title}
  {\bibinfo {title} {Critical quantum metrology with a finite-component quantum
  phase transition},\ }\href {https://doi.org/10.1103/PhysRevLett.124.120504}
  {\bibfield  {journal} {\bibinfo  {journal} {Phys. Rev. Lett.}\ }\textbf
  {\bibinfo {volume} {124}},\ \bibinfo {pages} {120504} (\bibinfo {year}
  {2020})}\BibitemShut {NoStop}%
\bibitem [{\citenamefont {Alushi}\ \emph {et~al.}(2025)\citenamefont {Alushi},
  \citenamefont {Coppo}, \citenamefont {Brosco}, \citenamefont {Di~Candia},\
  and\ \citenamefont {Felicetti}}]{alushi2025collective}%
  \BibitemOpen
  \bibfield  {author} {\bibinfo {author} {\bibfnamefont {U.}~\bibnamefont
  {Alushi}}, \bibinfo {author} {\bibfnamefont {A.}~\bibnamefont {Coppo}},
  \bibinfo {author} {\bibfnamefont {V.}~\bibnamefont {Brosco}}, \bibinfo
  {author} {\bibfnamefont {R.}~\bibnamefont {Di~Candia}},\ and\ \bibinfo
  {author} {\bibfnamefont {S.}~\bibnamefont {Felicetti}},\ }\bibfield  {title}
  {\bibinfo {title} {Collective quantum enhancement in critical quantum
  sensing},\ }\href@noop {} {\bibfield  {journal} {\bibinfo  {journal}
  {Communications Physics}\ }\textbf {\bibinfo {volume} {8}},\ \bibinfo {pages}
  {74} (\bibinfo {year} {2025})}\BibitemShut {NoStop}%
\bibitem [{\citenamefont {Wiseman}\ and\ \citenamefont
  {Milburn}(2009)}]{wiseman_milburn_2009}%
  \BibitemOpen
  \bibfield  {author} {\bibinfo {author} {\bibfnamefont {H.~M.}\ \bibnamefont
  {Wiseman}}\ and\ \bibinfo {author} {\bibfnamefont {G.~J.}\ \bibnamefont
  {Milburn}},\ }\href@noop {} {\emph {\bibinfo {title} {Quantum Measurement and
  Control}}}\ (\bibinfo  {publisher} {Cambridge University Press},\ \bibinfo
  {year} {2009})\BibitemShut {NoStop}%
\bibitem [{\citenamefont {Di~Candia}\ \emph {et~al.}(2023)\citenamefont
  {Di~Candia}, \citenamefont {Minganti}, \citenamefont {Petrovnin},
  \citenamefont {Paraoanu},\ and\ \citenamefont {Felicetti}}]{di2023critical}%
  \BibitemOpen
  \bibfield  {author} {\bibinfo {author} {\bibfnamefont {R.}~\bibnamefont
  {Di~Candia}}, \bibinfo {author} {\bibfnamefont {F.}~\bibnamefont {Minganti}},
  \bibinfo {author} {\bibfnamefont {K.}~\bibnamefont {Petrovnin}}, \bibinfo
  {author} {\bibfnamefont {G.}~\bibnamefont {Paraoanu}},\ and\ \bibinfo
  {author} {\bibfnamefont {S.}~\bibnamefont {Felicetti}},\ }\bibfield  {title}
  {\bibinfo {title} {Critical parametric quantum sensing},\ }\href
  {https://doi.org/10.1038/s41534-023-00690-z} {\bibfield  {journal} {\bibinfo
  {journal} {npj Quantum Information}\ }\textbf {\bibinfo {volume} {9}},\
  \bibinfo {pages} {23} (\bibinfo {year} {2023})}\BibitemShut {NoStop}%
\end{thebibliography}%

\end{document}